\documentclass[reprint, amsmath,amssymb, aps]{revtex4-1}
\usepackage{xcolor}
\usepackage{graphicx}

\usepackage[unicode,
            pdfauthor={Yannick Feld, Alexander K. Hartmann},
            pdftitle={Large-deviations of SIR model around the epidemic threshold},
            pdfsubject={SIR}]{hyperref}
\usepackage{grffile}

\def\equationautorefname~#1\null{Eq.\,(#1)\null}
\def\figureautorefname~#1\null{Fig.\,#1\null}
\def\sectionautorefname~#1\null{Sec.\,#1\null}

\begin{document}

\title{Large-deviations of the SIR model around the epidemic threshold}

\author{Yannick Feld}
\email{yannick.feld@uol.de}
\homepage{https://www.yfeld.de}
\author{Alexander K. Hartmann}
\affiliation{Institut f\"ur Physik, Carl von Ossietzky Universit\"at Oldenburg, 26111 Oldenburg, Germany}

\date{\today}

\begin{abstract}

  We numerically
  study the dynamics of the SIR disease model on small-world networks
  by using a large-deviation
  approach. This allows us to obtain the probability density function
  of the total fraction of infected nodes and of the maximum fraction of
  simultaneously infected nodes
  down to very small probability densities like $10^{-2500}$.
  We analyze the structure of the disease dynamics and observed
  three regimes in all probability density functions,
  which correspond to quick mild, quick extremely severe
  and sustained severe dynamical evolutions, respectively. Furthermore,
  the mathematical rate functions of the densities are investigated.
  The results indicate that the so called large-deviation property holds
  for the SIR model. Finally, we measured
  correlations with other quantities like
  the duration of an outbreak or the peak position of  the
  fraction of infections,
  also in the rare regions which are not accessible by standard simulation
  techniques.

\end{abstract}

\maketitle

\section{Introduction}

The modeling of the spread of epidemic diseases has always been
a central aspect in statistics, applied mathematics and statistical mechanics
\cite{hethcote2000,andersson2000,Pastor-Satorras_2015,walters2018,tang2020}.
Due to the present outbreak of the SARS-CoV-2 pandemic,
interest in this field has risen even more
\cite{dehning2020,roda2020,mandal2020,Liu_2020,chimmula2020,weissman2020,karaivanov2020}.
Disease spreading can be modeled in many different ways, e.g, with
ordinary differential equations like in the mean-field version of the
\emph{susceptible-infected-recovered}
(SIR) model \cite{Kermack1927} or with agent-based approaches
\cite{Eubank2004, Ferguson2005}. Also other fields are involved,
e.g., Bayesian analysis \cite{gelman2013}
to estimate model parameters \cite{dehning2020,Liu_2020}
or machine-learning approaches \cite{bishop2006}
to predict the future development of an outbreak
\cite{guo2017,ardabili2020,chimmula2020,kim2021}.
We refer to recent review articles for a good overview of the topic  
\cite{Wang2017, wang2019, wang2016_review}.

Given the large population of humans and animals on our planet and the high
number of active and potentially threatening viruses or bacteria,
the actual number of pandemic diseases is surprisingly small.
Thus, the outbreak of a \emph{specific}
pandemic is actually a \emph{rare event}, i.e.,
occurs, looking at each single type of disease, with a very small
probability. For example, the disease might be very active in one
population of, e.g., bats, but much rarer contacts or rare mutations
are needed to allow the transfer to another population, like humans,
maybe even requiring unknown intermediate animals.
Hence, it is natural to consider the application
of large-deviation approaches to study disease dynamics.
So far this was done only a few times, e.g.,
the large-deviation principle
was investigated analytically  \cite{pardoux2017} by
generalizing an approach of Ref.~\cite{dolgoarshinnykh2009}
 for simple mean-field epidemic models.

 A more realistic modeling of epidemic dynamics beyond
 mean-field level is generally obtained by
studying the dynamics on
networks \cite{Pastor-Satorras_2015}.
These networks represent the contacts between the individuals
or groups of individuals. 
This, of course, can become arbitrarily complex, e.g.,\ by combining several network layers, 
which can then represent different environments of contact \cite{Liu2018_reproduction}.
Depending on the structure of a network and
on the epidemic parameters, like transmission probability and
recovery probability, the infection of a single node might
stay contained or might lead to a pandemic outbreak. The critical
value of, e.g., the transmission probability, beyond which an pandemic outbreak
occurs,
i.e., a percolation of the infected nodes, is called the \emph{epidemic
  threshold}. For not too complex models,
the epidemic threshold of disease models can be analyzed
by using a variety of analytical methods, e.g., the mean-field method, its
quenched version, or dynamic message passing approaches.
\cite{Newman2002,Boguna2003,verdasca2005,Gross2006,gomez2010,Wang2016, Li2018}.
Naturally, for
more complex models it is even harder to obtain analytical results,
thus computer simulations \cite{practical_guide2015} are applied instead.

To our knowledge, for the study of disease spreading on networks
with respect to large-deviations and rare events
no results are available, let it be analytical or numerical.
Thus, to start to establish such approaches in the field
of disease dynamics, here the simple case of the
SIR model on standard networks
drawn from a small-world \cite{watts1998} ensemble is considered. This is motivated
by the fact that physical contact networks between humans
resemble small-world-like networks \cite{Eubank2004}.
However, the methods applied here can be used for all types of networks.
We apply large-deviation
techniques \cite{bucklew2004,Hartmann2002,Hartmann2014}
that are based on the Markov-chain approaches Wang Landau \cite{WangLandau-2001}
and entropic sampling \cite{Lee-1993}. In this way we are able to explore the
probability density function (pdf) of the fraction of infected nodes
 down to values as small as $10^{-115}$. For the
 pdf of the maximum of the fraction of simultaneously infected nodes,
 we reach values as
 small as $10^{-2500}$. For both quantities, we look at the
 respective mathematical
 \emph{rate functions}, to verify whether the \emph{large-deviation
   principle} holds
 \cite{denHollander2000,touchette2009,dembo2010,touchette2011}.
 This gives a complete
 description of these stochastic quantities, over the full range of the
 support of the distributions. First, this is desirable from a fundamental
 research point of view. Second,
we are able to investigate correlations between different quantities, e.g.,
how the fraction of infected  nodes  corresponds to how quickly the disease
dies out. By using a large-deviation technique, we are able
to study these correlations much beyond the typical behavior. Thus, we
can also analyze extremely severe as well as  extremely mild disease
progression, and try to identify their possible causes
through looking at their correlations. The much broader
understanding gained in this way could be one piece to
help to better prevent pandemic outbreaks in the future, in particular
if it is applied not to the general model but for a specific case
tailored to the epidemic under scrutiny, respectively.

The paper is organized as follows: First, the SIR model is introduced and
its dynamics and the main measurable quantities are defined. Next,
we define the ensemble of networks we use. In the main methodological section,
we present the algorithms used for
sampling the rare events
and how we have to set up the simulation of the SIR dynamics to embed it into
the large-deviation scheme.
Our results come in three parts. We begin by investigating
the ensemble with standard techniques to identify interesting points
in parameter space. For these points, large-deviation
simulations are performed to obtain the distributions of the total fraction $C$
of infected individuals and of the maximum fraction $M$
of simultaneously infected individuals,
respectively. We finish with a summary and an outlook.

\section{SIR model}

\label{sec-sir-model}
Let there be a given \emph{connected} network with $N$ nodes,
where the nodes represent
individuals and the edges contacts. The term \emph{connected} here
means that there is only one connected component, i.e., all nodes
can be reached from all other nodes through paths along edges.
Each node is in one of the three states  \emph{susceptible} (S),
\emph{infected} (I) or \emph{recovered} (R).

For any given configuration of states in the network,
at each time step a node can change its state as follows:
The probability of an infected node infecting a specific susceptible
neighbor is given by the
\emph{transmission probability} $\lambda = \text{const}$.
The probability of an infected node recovering in a given time step
is given by $\mu = \text{const}$. A node in the recovered state
remains recovered forevermore.

\begin{figure}[htb]
  \includegraphics[width=8.6cm]{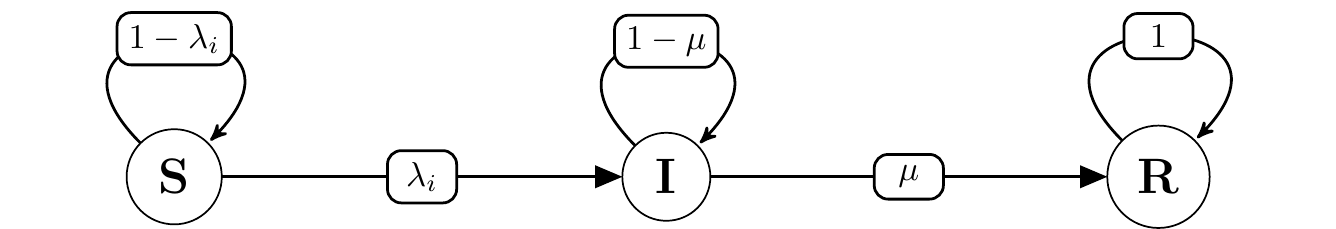}
  \caption{Showing transition probabilities for node $i$ at a given time step}
  \label{transition_sir_pr}
\end{figure}

We consider  a node $i$ in S state, which has
$A_i$ adjacent infected nodes.
Since  each infected neighbor has a probability of $\lambda$ to infect
node $i$, the probability for node $i$ to become infected in a time step is
\begin{equation}
  \lambda_i  = 1 - \left(1-\lambda \right)^{A_i}\,.
\label{eq:infection}
\end{equation}
All possible transitions between the states of a node
are shown in \autoref{transition_sir_pr}.

For all disease dynamics we consider, as initial state at discrete time $\tau=0$,
one particular node (in the following called node 0)
set to the infected state, while
all other nodes are susceptible. Our simulations
\cite{practical_guide2015}
are performed at discrete times $\tau \to \tau+1$
by applying the above mentioned rules in a parallel fashion
to all nodes. This is repeated  until the disease dies out,
i.e., no infected nodes remain, or if a maximum chosen time is reached.
Such a
development we call the \emph{time evolution of an outbreak}
from here on out. Note that the outbreak might be very small, with just
node 0 being initially infected and recovering after some time before
any other node is infected. Clearly, unless $\lambda \ll \mu$, this
will not occur too often.

Due to the probabilistic nature of the problem, 
multiple outbreak simulations will generally lead to different results.

To describe the time evolution of an outbreak,
let us introduce a few quantities:
Let $s(\tau)$, $i(\tau)$ and $r(\tau)$ be the fractions of susceptible, infected and recovered nodes at time step $\tau$ respectively.
Let $c(\tau) = i(\tau) + r(\tau)$ be the fraction of the total, i.e.,
cumulative infections, which have occurred up to time step $\tau$.
These quantities depend on the time step. 
To describe the global characteristics of an outbreak,
the following two quantities are introduced:
\begin{equation}
  C =\max_\tau(c(\tau)) \equiv c(\infty)
\end{equation}
describes the fraction of the network that caught the disease during the
outbreak and is  therefore a measure for its severity. This is the standard
quantity to distinguish between a local outbreak and a pandemic.
\begin{equation}
  M = \max_\tau(i(\tau))
  \end{equation}
denotes  the peak fraction of
nodes that happen to simultaneously be
in the infected state during an
outbreak and is, therefore, a relevant quantity for the health care system.

\section{Ensemble}

In our work, we investigate a small-world network ensemble
\cite{watts1998, amaral2000, barrat2001, watts2004},
because contact networks between individuals are highly
connected small-world-like networks \cite{Eubank2004}.
We use the same implementation as we used in Ref.\ \cite{feld_2019}.

The network is constructed as follows: Let there be $N$ nodes $i=0,\dots,N-1$.
First, the nodes are arranged in a ring structure, meaning every node is
connected
to its next and second next neighbor by the edges $\{i, i+1\}$ and
$\{i, i+2\}$
(nodes $N$ and $N+1$ are identified with index 0 and 1 respectively).

To gain small-world characteristics, next some of the edges created in the first step are made
\emph{long-range}, i.e.,
each edge will be rewired with probability $p$.
To rewire an edge $\{i,j\}$, with $j=i+1$ or $j=i+2$,
a random node $j^\prime \neq i$ is drawn and the edge
is changed to $\{i,j^\prime\}$.
Throughout this paper, $p=0.1$ is used.

As mentioned, we only consider \emph{connected} networks,
i.e., networks where there exists a path of edges between any two nodes.
We use depth first search to verify, whether any created
network is connected or not.
If the network is not connected, the whole network is discarded and
the construction process repeated, until a connected network is generated.

\section{Algorithms}
\label{sec_disease_spreading}

The straightforward way to perform the outbreak simulations
outlined in Sec.~\ref{sec-sir-model},
often called \emph{simple sampling}, is to start with the initial
state and, while performing the iterations, draw the necessary
random numbers independently ``on demand''. This will generate
typical outbreaks, i.e., when performing $K$ independent runs, one
can efficiently sample events
which occur with probabilities not smaller than
$O(1/K)$.

We are interested in the large-deviation properties of the outbreaks,
i.e., want to access events that occur with much smaller probabilities.
To achieve this, we have
to control the dynamics of the outbreaks. This works by biasing them
in a suitable way within a Markov chain Monte Carlo (MCMC)
simulation \cite{Hartmann2014},
as explained in Sec.~\ref{sec:rare-event}. But in order to use
the outbreak simulation as the basic element
within an MCMC simulation, we have to make
it accessible for control, as explained in the next section.

All large-deviation simulations are
for a fixed given network. As stated below, we average only over
few networks, or, for a large number $N$ of nodes, only
one given network is considered due to assumed self-averaging. Thus, we are \emph{not}
interested in rare properties induced by rare network structures.
This is justified, because the contact network of a population
of individuals is usually given.
Thus, what we are interested in are typical and rare dynamical
processes taking place on typical networks.

\subsection{Outbreak simulation\label{ld_energy}}

To analyze this model with large-deviation methods down to very
small probabilities, we need a way to manipulate the randomness of the
spread of disease in a controlled fashion.
This is done by manipulating the random numbers utilized within the simulation.

An easy way to achieve this, is to draw the random numbers beforehand,
store them in one or several vectors \cite{crooks2001,Hartmann2014} and pick
numbers from the vector whenever needed.
That means, an educated guess is required about how many time steps
$t_{\max}$ are
needed  for the simulation in order to make the vectors large enough.
Clearly, the choice
of $t_{\max}$ will depend on the values of $\lambda$ and $\mu$ and
will be determined below.

Now, the random numbers to be drawn beforehand are contained in two
arrays $\xi_\lambda[l]$ and $\xi_\mu[l]$ with $l=0,1,\ldots t_{\max} N$.
The entries shall be drawn uniformly between 0 and 1 each. The MCMC
approach will manipulate these two vectors in order to control the
outbreak simulation.
The basic assumption used in the MCMC approach is that the state
of a system, here the entire evolution of an outbreak, changes only slightly, if the random numbers
are changed only slightly. For this purpose each random number will be
assigned a specific purpose or use. This implies that
any random number can occasionally be ignored.

Now the use of the random numbers in one outbreak simulation is detailed.
Let $\tau\ge 0$ be the current time step. To calculate the states
of the nodes for the next iteration, we
first iterate over all susceptible nodes $i$, that have at least
one infected neighbor.
The probability for $i$ to be infected is
$\lambda_i$ as shown in \autoref{eq:infection}.
To decide whether the node should be flagged for becoming infected
at time $\tau+1$
 the random number $\xi_{\lambda}[\tau N + i]$ is used, i.e.,
it will be flagged to become infected if $\xi_{\lambda}[\tau N + i]< \lambda_i$.
Of course this means that for all nodes $i$ which have no infected neighbors, the corresponding
entries of $\xi_{\lambda}$ are ignored.

Next, we iterate over all
infected nodes $i$.
We use the random number stored at  $\xi_{\mu}[\tau N + i]$,
to flag the state of node $i$ to be recovered in the next time step,
which occurs with probability $\mu$.

Afterwards all nodes that are currently flagged to become infected are set to infected.

Note that, technically, one could store all needed random numbers
from $\xi_{\lambda}$ and $\xi_{\mu}$ in one single array. We found
this splitting more convenient, in particular because it allows to easily manipulate
the arrays in different ways.

The underlying network is not changed during the simulation, it represents
a typical society. Furthermore, because the actual outbreak 
always starts with only node 0 being infected,
no other randomness is present except the one 
contained in the two vectors of random numbers. Thus, the dynamic evolution
and any measurable quantity are deterministic functions
$f(\xi_{\lambda}, \xi_{\mu})$.

\subsection{Large-deviation Sampling \label{sec:rare-event}}

Our goal is to calculate the probability density function (pdf) $P(E)$
for a given network $\mathcal{G}$ and given values of $\lambda$ and $\mu$.
Here, $E$ stands for a measurable quantity of the spread of disease, in our case either $E\equiv C$ or $E\equiv M$.
In the following, $E$ will be referred to as \emph{energy}.

To calculate the pdf over a large range of the support, possibly over its full
support, one usually must be able to obtain it in the region
of very small probability densities as well. To achieve this within numerical
simulations \cite{practical_guide2015},
specific large-deviation algorithms \cite{bucklew2004} can be applied.
Such approaches have been used to study various equilibrium
and non-equilibrium problems like alignment scores of protein sequences
\cite{Hartmann2002,align_book,align_long2007}, nucleation \cite{adams2010},
properties of random networks \cite{rare-graphs2004,largest-2011,diameter2018},
dynamics of the totally
asymmetric exclusion process \cite{giardina2006,lecomte2007},
traffic models \cite{nagel_schreckenberg2019},
calculation of partition functions \cite{partition2005},
dynamics of model glasses \cite{elmatad2010},
dynamics of Ising ferromagnets
\cite{jack2010,Hartmann2014}, statistics of negative-weight
percolation \cite{nwp_MK_imp_samp2013} and RNA work processes \cite{werner2021}.

Various large-deviation algorithms exist. Here, we applied
an approach based on the Wang-Landau (WL) algorithm \cite{WangLandau-2001}.
Although the general approach is well known, we present the main steps
along with the details that are necessary to reproduce our results.

The algorithm starts with a non-normalized estimate $g(E)$ of the density
of states  for the energy $E$. In case one does not have any prior
information, like here,
one starts with an unbiased estimate $g(E) = 1\,\forall E$.
The algorithm will  iteratively refine $g(E)$ to converge
closely to the true pdf.
This is achieved by creating a
Markov chain in the space of all possible outbreaks for a given network and given
initial state S, I or R of each node.
Since, as shown in the previous section, each
outbreak has a one-to-one correspondence to the two arrays
$\xi_\lambda$ and $\xi_\mu$, the Markov chain is actually performed
in the space of all possible assignments of random number entries from
$[0,1]$ to these two arrays.
We denote by $(\xi^n_\lambda, \xi^n_\mu)$ the current
configuration at Markov step $n$.
For each of such a configuration, a full outbreak simulation is 
performed and the energy, i.e., the cumulative or peak fraction
of infections, is read
off. Thus, as mentioned, this energy is just a deterministic
function of the configuration:
$E_n = E\left(\xi^n_\lambda, \xi^n_\mu\right)$.

To perform the Markov chain, we use in particular the
Metropolis-Hastings MCMC method \cite{metropolis1953,hastings1970,newman1999}.
Therefore,
each step in the Markov chain consist of generating a \emph{trial
  configuration} $(\tilde \xi_\lambda, \tilde \xi_\mu)$ from the current
configuration, which will be accepted or rejected, as detailed more below.

First, we explain how the trial configurations are generated here.
A combination of three different possible moves is used. 
The moves are
all based on the current configuration, i.e., they start with
$(\tilde \xi_\lambda, \tilde \xi_\mu)=(\xi^n_\lambda, \xi^n_\mu)$.
One of the
three following change operations is randomly selected:

With a high probability of 98\%, we just
perform \emph{random changes} as follows:
We randomly choose one of the  two arrays $\tilde \xi \in
\{\tilde \xi_\lambda, \tilde \xi_\mu\}$,
draw a random index $k$ and a random number $\chi \in [0,1]$ uniformly
and set $\xi[k] = \chi$. This is repeated $B$ times.
As a rule of thumb, B should be chosen such that about 50\% of the trial configurations are accepted,
which is what we aim for.
The actual numbers are stated in the results section.
Note that the correctness of the method does not depend on the acceptance rate,
however, it does affect efficiency.
It is clear that this move alone can reach all possible
configurations of $(\xi_\lambda, \xi_\mu)$, which
means that ergodicity is fulfilled. Nevertheless, for a better
convergence, we include two more moves:

With a probability of 1\%, we perform a \emph{rotation}, i.e.,
$\tilde \xi_\lambda$ and $\tilde \xi_\mu$ are rotated by
$N$ elements to the left or to the right, with periodic boundaries.
This roughly corresponds to
shifting the  resulting time series of the outbreak by one time
step to the left or right.

Also with a  1\% probability we perform a \emph{swap}. Here,
we draw two random indices $\iota$ and $\nu$
and swap the values $\xi_\lambda [\iota] \leftrightarrow \xi_\lambda [\nu]$ and
$\xi_\mu [\iota] \leftrightarrow \xi_\mu [\nu]$.
This is repeated $B$ times to create one trial configuration.

Note that these moves do
not skew the probability of
the resulting random-number vectors  in any direction, since all entries
are always uniformly drawn from $[0,1]$.

For the trial configuration of random numbers a complete outbreak
simulation has to be performed again, resulting in the corresponding energy
$\tilde{E} = E(\tilde{\xi}_\lambda, \tilde{\xi}_\mu)$.

The trial configuration will now be accepted, i.e.,
$(\xi^{n+1}_\lambda,\xi^{n+1}_\mu)=
(\tilde{\xi}_\lambda, \tilde{\xi}_\mu)$
and therefore $E_{n+1} = \tilde{E}$, with a Metropolis-Hastings probability
\begin{equation}
  p_{\text{acc}} = \min \left(1, \frac{g(E_n)}{g(\tilde{E})}\right)\,.
  \label{acc_prob}
\end{equation}
If the trial configuration is rejected, the current
configuration is kept, i.e., $(\xi^{n+1}_\lambda,\xi^{n+1}_\mu)=
(\xi^{n}_\lambda,\xi^{n}_\mu)$  and therefore $E_{n+1} = E_n$.

As usual for the WL algorithm, next the density estimate $g$ is updated
using a multiplicative factor $f>1$, i.e., $g\left(E_{n+1}\right) \to f
g\left(E_{n+1}\right)$, while for all other values of $E$, $g(E)$
remains the same.
One can start with a rather large factor
like $f=e\approx 2.71$. The factor is then
reduced towards $1$ during the simulation.

However the saturation of the final error becomes a problem for the original
WL algorithm (\cite{Yan-2003}, see also
Refs.~\cite{Belardinelli-2007, Belardinelli-2008, Belardinelli-2016}).
The algorithm introduced by Belardinelli and Pereyra \cite{Belardinelli-2007}
is used to circumvent the problem, since it was shown \cite{Belardinelli-2016},
that error saturation does not become a problem for this alternative algorithm.
The main difference between this algorithm and the original WL is in how
the factor $f$ is updated during the simulation, for details see
the citations.

Still, the WL algorithm and its variants do not fulfill detailed balance.
Therefore we perform entropic sampling \cite{Lee-1993} afterwards.
We start with the estimate $g(E)$ as computed by WL.
Entropic sampling is very similar to WL. The same method is used to
generate a Markov chain and accept the states based on the probability
\autoref{acc_prob}.
This time, however, we do not update $g$, but instead maintain a histogram
$H(E)$ of visited states.
We always employ entropic sampling for the same number of steps, as
were used for the preceding Wang-Landau runs, respectively.

To finish the entropic sampling simulation, the desired pdf can be
calculated. First, a non-normalized pdf is calculated
\begin{equation}
  \tilde{P}\left(E\right) =g(E)  H(E)
\end{equation}
for all bins, where $H(E)>0$. For all other bins the pdf would be unknown.
Then the pdf is normalized
\begin{equation}
  P(E) = \frac{\tilde{P}\left(E\right)}{\int_{\widehat{E}}
    \tilde{P}(\widehat{E}) d\widehat{E}}\,.
\end{equation}

During the simulation using the entropic sampling, we occasionally sample,
i.e. store
trajectories of outbreaks, which can be analyzed later on. This will
lead to a rather uniform sampling of the trajectories
with respect to the measured energy, $C$ or $M$.

Calculating the pdf over the whole regime at once can be rather challenging.
To make it more feasible, the $E$ range is split in multiple
overlapping intervals \cite{Schulz2003, landau2004}.
For each of those intervals we performed a WL and an entropic
sampling simulation.
Finally the resulting pdf are merged to obtain a full pdf.
This can be done, because the pdfs of the overlapping regions have to
match, at least within statistical
fluctuations \cite{WangLandau-2001,Hartmann2002}.

For some of these overlapping intervals we had problems with the ergodicity,
which can be observed if not the full interval is visited, or if the
distribution from neighboring intervals does not match well.
To circumvent the problem, we use a replica exchange Wang-Landau (REWL)
algorithm \cite{vogel_2014_0, vogel2014_1, Li_2014} for the affected pdfs,
which works similar to the Wang-Landau algorithm described above,
but regularly attempts to exchange configurations
between independent simulations on different intervals, utilizing a suitable
Metropolis criterion. Again, we refer to the literature for details.
Note that we also applied the replica exchange approach to the entropic
sampling to obtain
the final pdf estimates.

\section{Simple Sampling Simulations}

To choose  points of interest in parameter space and a suitable
length $t_{\max}$ of the outbreak simulations, we have performed
 some test simulations prior
to the large-deviation simulations.

\subsection{Critical transmission probability}

\begin{figure}[htb]
  \includegraphics{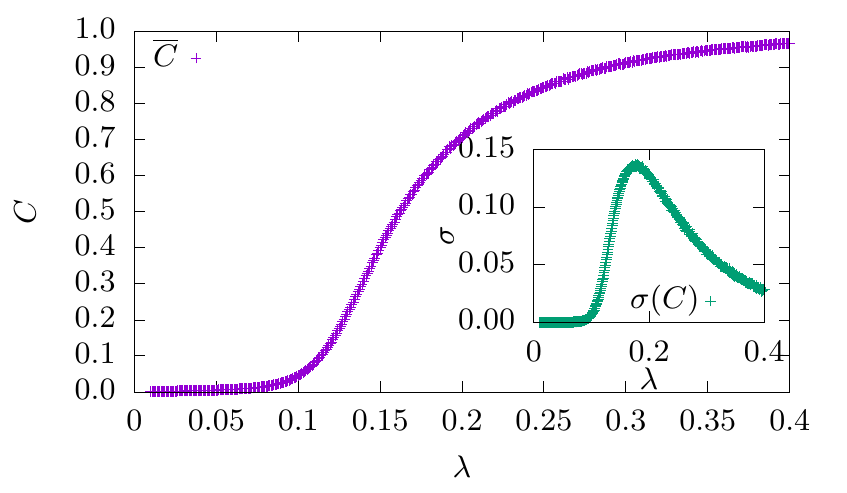}
  \caption{(color online)
Example for the average $\overline{C}$ as a function of the    transmission
    probability $\lambda$ for $N=800$ and $\mu=0.14$.
    The inset shows the variance. Error bars are smaller than symbol sizes.}
  \label{example_trans}
\end{figure}

We want to analyze the behavior of the model in the non-pandemic phase,
in the pandemic phase, and close to the epidemic threshold.
Since we work in discrete time, the parameters are not rates
but probabilities.
Thus, unlike to the continuous-time case, there is no natural or neutral
time scale and we cannot set one of the probabilities to 1.
Therefore, the recovery probability is chosen to be 
$\mu=0.14$
as a working basis in all simulations.
Since we are still free to choose $\lambda$, the general
results should not depend much on the specific value of $\mu$, unless it approaches
0 or 1.
Thus, the task is to determine
the critical transmission probability $\lambda_c$ of
the epidemic threshold. For this purpose,
we investigate  network sizes up to $N=3200$.
For each value of $N$, we generated 200000 random networks  
and performed outbreak simulations
for each network. Initially, only node 0 is infected,
while all other nodes are
susceptible. Here, all outbreaks are iterated until no infected node
remains.
We measure the average cumulative fraction $\overline{C}(\lambda)$
of infected nodes and also calculated the variance $\sigma(C)$ for each combination
$(N, \lambda)$.
Errors are estimated with bootstrap resampling
\cite{efron1979}.
For a different number $N$ of nodes
the curves $C(\lambda)$  change so little, that it would be hardly
visible. We therefore only show an example of this in  \autoref{example_trans}.

\begin{figure}[htb]
\includegraphics{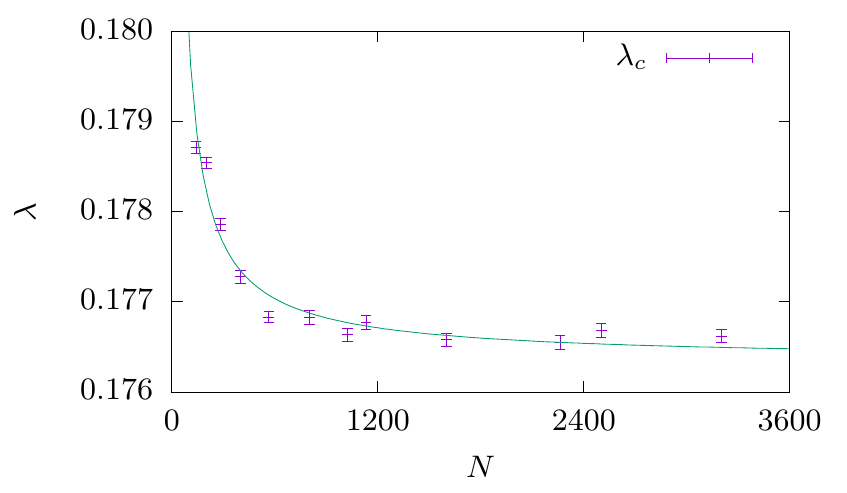}
\caption{(color online)
Critical transmission  $\lambda_{c}$ for $\mu=0.14$ as a function of
  the number $N$ of nodes with fit to $\lambda_c(N)$ from
  \autoref{eq:fit:crit}.}
\label{critical_trans}
\end{figure}

We define the finite-size critical transmission $\lambda_c(N)$
as the peak of the variance  $\sigma(C, N)$.
To measure the peak of the critical transmission, we fit  Gaussian-shaped
functions
around the maxima, respectively.
We then apply standard finite-size scaling to calculate the
critical transmission rate $\lambda_c(\infty)$
by fitting

\begin{align}
  \lambda_c(N) = \lambda_c(\infty) + a  N^{-b}
\label{eq:fit:crit}
\end{align}
to the data, as shown in \autoref{critical_trans}. 

We obtain a value $\lambda_{c}(\infty) = 0.1763(2)$ for the
critical transmission.
The other fit parameter were $a = 0.24(24)$ and $b=0.91(21)$, with rather
large error bars, but these
values are not of interest here. Note that fitting 
$\lambda_c(N) = \lambda_c(\infty) + a_2 \log(N)^{b_2}$ looks very similar 
and leads to a similar critical transmission $\lambda_c(\infty) = 0.1750(4)$. 

\subsection{Disease duration}
\label{sec_disease_duration}

For the large-deviation simulation, we cannot simply run each outbreak
simulation until the disease dies out, because the MCMC scheme operates
with a vector of random numbers which must be of fixed length. Thus,
we have to find a suitable time scale for the duration of the
outbreak simulations.

For this purpose, we performed simulations in the same manner as described
in the previous section and measured the
duration $\Delta t$ it takes until no infected
nodes remain, i.e., $i(\Delta t) = 0$, for each simulation.
After this time, the state of the nodes
will not change because the outbreak dynamics are finished.
For each parameter set $(N, \lambda)$, we measured 100000 randomly generated networks.

\begin{figure}[htb]
  \includegraphics{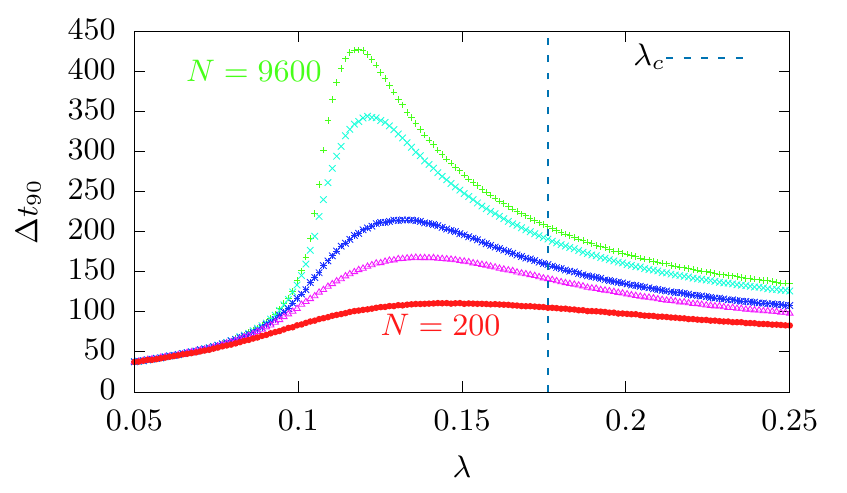}
  \caption{(color online) The duration $\Delta t_{90}$ until the outbreaks
    are completely finished for 90\% of the independent outbreak simulations
    as a function of
    $\lambda$, for different system sizes $N$. The largest and smallest value of $N$ are labeled.
    In between the values behave monotonously. The used values are $N \in \{200,600,1200,4800,9600\}$.
    The dashed line indicates the value of $\lambda_c$.}
  \label{percent90Curve}
\end{figure}

For each considered parameter set $(N,\lambda)$ the characteristic 
time $\Delta t_{p}$, which describes
how long it takes until p\% of the outbreak dynamics are finished, is calculated.
As example we show the measured curves for 
$\Delta t_{90}$ in \autoref{percent90Curve}.
The curves show the typical signs of a dynamical phase transition accompanied by critical slowing down \cite{nazarimehr2020}.
Critical slowing down plays an important role for \emph{early warning signals} of infectious disease transitions \cite{southall2021}.
Interestingly, the outbreak takes longest well below the epidemic threshold 
$\lambda_c$, which fits with previous observations \cite{holme2013}.

\begin{figure}[htb]
  \includegraphics{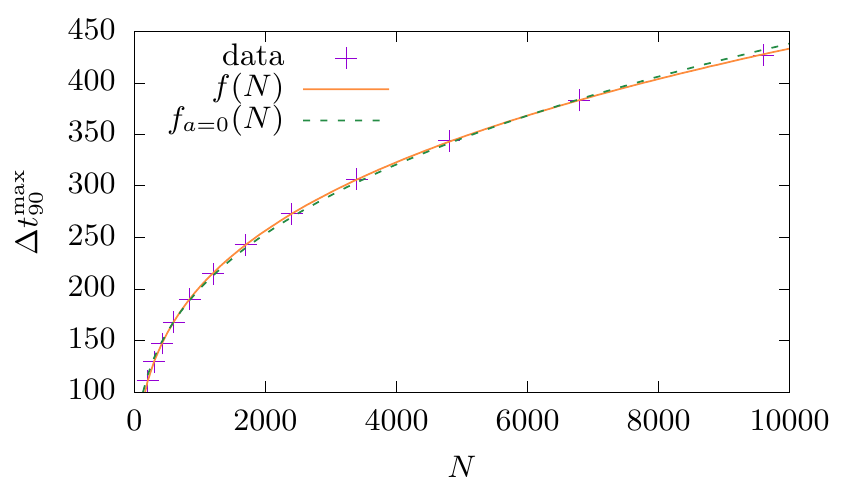}
  \caption{(color online) Maximum $\Delta t_{90}^{\max}$ over all 
values of $\lambda$  of how long it takes until 90\% of outbreaks are finished,
     as a function of $N$.    The continuous line shows the fit to  $f(N)$, while the dashed line 
     shows the fit to $f_{a=0}(N)$.}
  \label{Delta_Max}
\end{figure}

To investigate the worst-case scenario, we look at
\begin{equation}
  \Delta t_{p}^{\max}(N) := \max_\lambda\left[\Delta t_{p}(N, \lambda)\right]\,.
\end{equation}
The result for $\Delta t_{90}^{\max}(N)$ can be found in \autoref{Delta_Max}. We used this result to
set up the length of the outbreak simulations within the large-deviation approach, see below. 
But beyond this technical aspect, it is also interesting
to investigate the scaling behavior. 
Critical slowing down often leads to a power-law behavior of the correlation length \cite{fisher1986, le2005}. 
Since $\Delta t_{90}^{\max}(N)$ can be understood as correlation length we fitted a power-law
\begin{equation}
  f(N) = a + b N^{c}
\end{equation}
to the $\Delta t_{90}^{\max}(N)$
data. This seems to describe the relation very well.
The fit parameter are $a=-50(4)$, $b=36(2)$ and  $c=0.2824(43)$. 
That means that the time it takes until 90\% of the outbreaks are over
scales roughly with the 4th root of the system size.

The fit still works well with a fixed $a=0$, i.e., for 
\begin{equation}
f_{a=0}(N) = b N^{c}
\end{equation}
which leads to $b = 19.1(6)$ and $c= 0.340(4)$.
Also note that both fits work exceptionally well for all  $\Delta t_{p}$ except for $p$ very close to $p=100\%$ or $p=0\%$.
It is even possible to find simple functions for the fitting parameter, 
i.e., $a(p)$ et cetera.
The functions which are obtained  by using these functional
 parameters, e.g., $f(N|p)=a(p)+b(p)N^{c(p)}$, also fit the data reasonably well.

Still, as this is only preparation for our large deviation sampling, we do not pursue this any further.

\begin{figure}[htb]
  \includegraphics{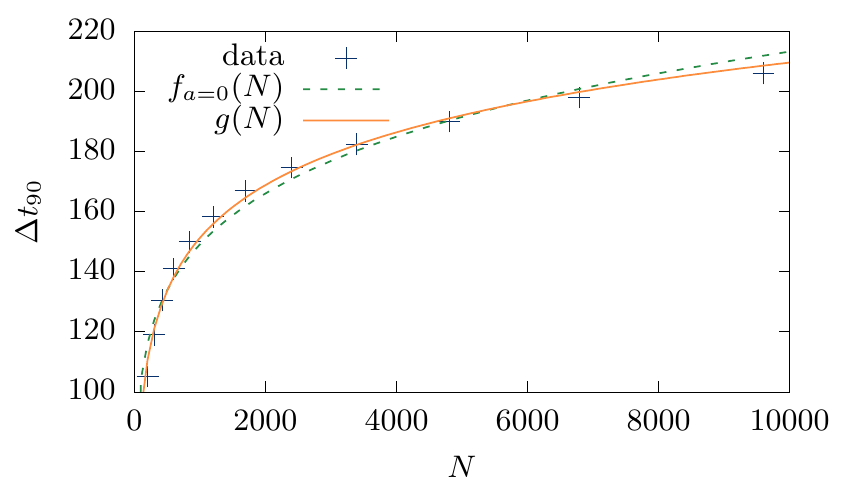}
  \caption{(color online) The time $\Delta t_{90}$ how long it takes until 90\%
    of outbreaks dynamics are finished as a function of the network size $N$
    for the critical value $\lambda=0.1763$.
    The lines show the results of fits to $f_{a=0}(N)$ and $g(N)$.}
  \label{delta_crit}
\end{figure}

It is also interesting to look at the duration right
at the critical transmission.
This can be found in \autoref{delta_crit}.
Here we set $a=0$ for the fit because the errors become
unreasonably large otherwise.
The obtained parameters from the fit  are
$b=51(3)$
and $c=0.1555(8)$,
which means that here the duration scales roughly only
with the 7th root of the system size.

Note that usually a power law with a small exponent cannot be well distinguished
from a logarithmic behavior. Therefore, we also fitted a logarithmic function
\begin{equation}
  g(N) = \alpha  \log(N \beta)\,.
\end{equation}
The quality of this fit is not good for the $\Delta t_{90}^{\max}$
data,  but works well for specific values of $\lambda$,
e.g., $\lambda = \lambda_c$ and is therefore included in
\autoref{delta_crit}. In the latter case
the obtained fit parameters are
$\alpha = 25.3(6)$
and $\beta = 0.39(6)$.

\section{Cumulative fraction $C$ of infections}

Using the large-deviation approach, we
now present the result for the distribution of the fraction
$C$ of cumulative
infected nodes. Note that the study of the large deviation refers to the
dynamics on a given network. Since in real-world situation the contact network is
given, we do not study rare-events with respect to rare networks here.
First, we present results for the pdf of $C$. They are obtained
by using Wang-Landau,
plus afterwards refining the result
with entropic sampling. If necessary, i.e., in case
we observed non-convergence, we applied REWL instead.

The parameters we use for the simulations are presented
in Tab.~\ref{tab:paramC}, for the
different networks sizes $N$ and values of the transmission probability $\lambda$.
At the critical point $\lambda=\lambda_c$, we study also finite-size
scaling by
considering different network sizes. For the representative values smaller and
larger than  the critical transition, $\lambda=0.1$ and $\lambda=0.4$,
respectively, we perform
simulations only for a rather large system size of $N=3200$. 

\begin{table}[h!]
  \begin{tabular}{|ll|llll|}
  $\lambda$ & N    & approach & $\#N$ & $\#I$ & 	$B$	 \\ \hline
  0.1763    & 200  & WL         & 15                & 24     &	75	\\
  0.1763    & 400  & WL         & 7                 & 24     &	150  \\
  0.1763    & 800  & WL         & 4                 & 24     &	166	\\
  0.1763    & 1600 & WL         & 1                 & 24     &	900	\\
  0.1763    & 3200 & WL         & 1                 & 24     &	2048	\\
  0.1763    &  6400 & WL         & 1                 & 48     &	3072	\\
  0.1       & 3200 & WL         & 1                 & 24     &	1024	\\
  0.4       & 3200 & REWL       & 1                 & 27     &	256	\\
  \end{tabular}
  \caption{\label{tab:paramC}
    Parameters for the simulations: transmission probability $\lambda$,
    number of nodes $N$, the approach used, the number $\#N$ of
    independent network realizations, the number $\#I$ of intervals used
    in the WL or REWL sampling  and the number $B$ of exchanges
    performed per MCMC attempt for the
    arrays $\xi_\lambda$ and $\xi_{\mu}$ of random numbers.}
  \end{table}

Note that we use
a recovery probability of $\mu=0.14$ everywhere,
which, as a side note, is in the range of recovery probabilities used to 
model the current corona virus pandemic \cite{karaivanov2020, dehning2020},
although there is a wide variablilty of models leading
to other parameter values 
\cite{Liu2021_china, Liu2021_closure, Pani2020, Parker2021}.

For small sizes,
we perform the full large-deviation sampling for a small number $\#N$
of few independently drawn networks, while for the largest sizes, where we
assume some kind of self-averaging, we study only $\#N=1$ generated
network. The latter case also corresponds somehow to the real-life
simulation, where only one contact network is given, but the dynamics
evolves randomly.

For the large-deviation simulation, we need to choose a length
of the
vector of random numbers, which determines the maximum time duration $t_{\max}$
of an outbreak
that can be covered. In theory, arbitrary long outbreaks are possible,
so one has to choose a cutoff time anyway. We have chosen as maximum
outbreak time of $t_{\max} = 3 \Delta t_{90}^{\max}(N)$, the latter one
as determined in \autoref{sec_disease_duration}.
To verify whether this is long enough,  we keep track,
during the large-deviation sampling, of
how often
the outbreak was \emph{un}finished after the given time.
When considering all the different network sizes,
transmission probabilities
and intervals of $E\equiv C$,
the highest, i.e., worst-case
frequency $f_{\neq}$ of observing a non-finished outbreak  occurred
for $N=3200$ in the interval of $E\in [0.47,0.56]$ with
$f_{\neq} \approx 1.4 \times 10^{-5}$.
Typically, the frequency was much lower, e.g., the worst interval for
$N=6400$ exhibited
$f_{\neq}\approx 1.7 \times 10^{-8}$. Since an unfinished outbreak
constitutes only few infected nodes anyway, this shows that in
order to observe extremely unlikely events in terms of $C$, and clearly $M$
anyway, one does not have to cover extreme unlikely long durations of
outbreaks and the choice of $t_{\max}$ is sufficient.

\begin{figure}[htb]
    \includegraphics{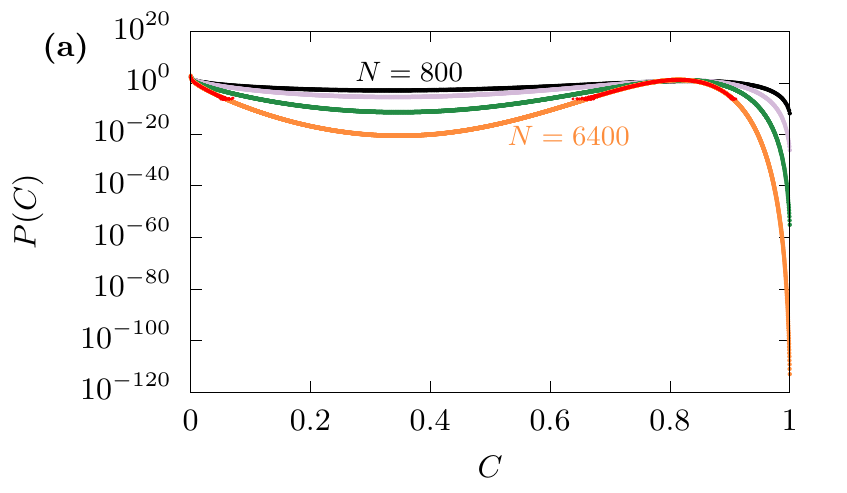}\\
    \includegraphics[width=0.49\linewidth]{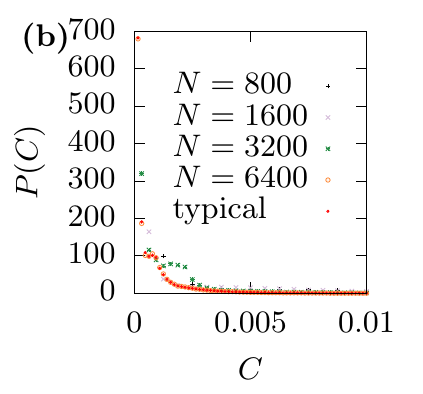}
    \includegraphics[width=0.49\linewidth]{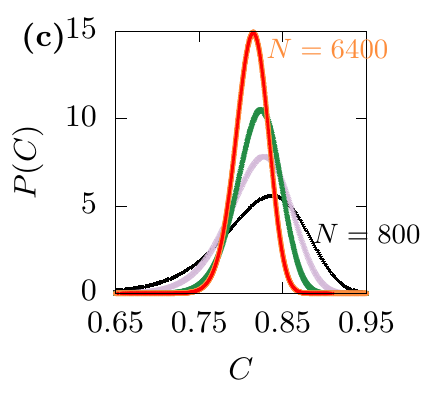}
    \caption{(color online) Probability density of total infections $C$ 
for $\mu = 0.14$ and $\lambda=0.1763$. (a) shows the full distribution
in logarithmic scale, while (b) and (c) highlight the two peak regions
in linear scale.  For (a) and (c) the largest and smallest values of 
$N$ are labeled,  in between the values behave monotonically,
except for a small area around $C=0.8$, where the order is  reversed, see (c). 
The used values were $N\in\{800,1600,3200,6400\}$.
We also included the typical-event sampling results for $N=6400$ in red.}
    \label{sw_multi_N}
\end{figure}

In \autoref{sw_multi_N} the probability density $P(C)$ is plotted for
different system sizes $N$.
Note that here and in the following the pdfs $P(E)$, 
where $E$ can be either of $C$ and $M$,  are always normalized
such that $\int_E P(E) = 1$.
Note also that we sample the histograms with the highest possible
resolution of one bin per possible value of $C$.
Whenever we average over different networks,
we calculate the pdfs for each of them and then merge them by
averaging the logarithmic probabilities and normalizing again.

We are able to measure the probability density
over the whole range of its support, extending over up to 115 decades in probability.
To put that into perspective, for $N=6400$ we calculated $C$
about $3.1 \times 10^9$ times, once per MCMC attempt, during
entropic sampling and WL combined. That means, if we use 
typical-event sampling to
create a histogram and estimate the probability density function with
the same numerical effort,
we are only able to resolve probabilities with a resolution of
about $10^{-9}$.
This is also shown in \autoref{sw_multi_N}, where we also see that the 
typical-event sampling results 
and the large deviation results agree very well.
To resolve the whole density function with typical-event sampling, one would need
about $3 \times 10^{105}$ times
as much computational power as used here.
The computational advantage would grow even higher for larger values of $N$.

Having the whole probability density function is interesting 
from an insurance perspective. 
Let's assume there is a cost function $\text{Cost}(C)$.
Using the pdf we measured one can easily calculate the exact expected value 
of the cost function. Even the very improbable cases will be relevant here 
because they will likely be associated with very high costs, e.g.,
the financial loss due to a pandemic.

As visible by the two peaks in the pdfs, the disease
either dies out very quickly, corresponding to the peak near
$C=1/N\approx 0$,
or about 80\% will contract the disease over the evolution of
the outbreak.
Intuitively this makes sense, as only one node is infected
in the beginning, thus
the disease dies out if that node recovers before infecting anyone.
If the disease does not die out quickly on the other hand, it will
persist until
a good fraction of the network is immune.
The observed behavior becomes more pronounced for larger networks,
as visible by a decrease of $P(C)$ for intermediate values of $C$.

\begin{figure}[htb]
  \includegraphics{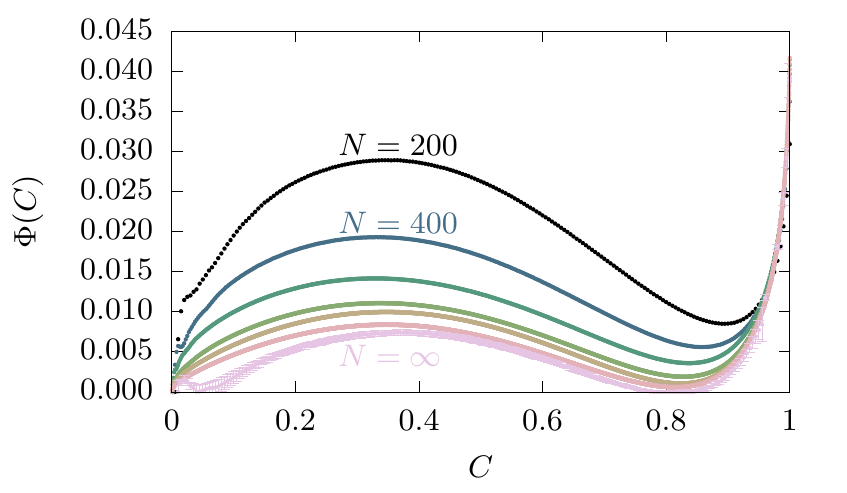}
  \caption{(color online)
Rate function $\Phi$ as a function of total infections $C$ for different values of $N$.
  The smallest two and the biggest value of $N$ are labeled. In between, the values behave
  monotonically. The values $N \in \{200,400,800,1600,3200,6400,\infty\}$ are used.
  $N=\infty$ is gained by extrapolation and all other values are measured.}
  \label{rate_fn}
\end{figure}

To relate to mathematical large-deviation theory, we also study
the empirical rate functions, defined as

\begin{align}
  \Phi \left(C, N \right) := -\frac{\ln(P(C))}{N} + \Phi_0
\end{align}
where $\Phi_0=\text{const}=\min_C \left(-\frac{\ln(P(C))}{N} \right)$,
such that the minimum of the rate function occurs at $\Phi=0$.
The calculated rate functions are displayed in \autoref{rate_fn}.

Since an apparent convergence is visible when increasing network size $N$,
 we also estimate the rate function for $N=\infty$ with
finite-size scaling, in a similar way to the
disease duration in  \autoref{sec_disease_duration}.
For this purpose we fit the function
\begin{equation}
  k(N) := \Phi_\infty + \eta N^{-\kappa}
\end{equation}
for each value of $C$.

The errors of the fit are used
as error bars and the result is included in \autoref{rate_fn}.
The results show that the numerically obtained rate function
seems to converge well. This means that the mathematical
\emph{large-deviation principle holds}, i.e., the size dependence on $N$
is in leading order given by
$P(C)\sim \exp\left(-N \Phi(C)+{\rm o}(N)\right)$. As a consequence
of this ``well behaving'', analytical progress regarding $P(C)$ might
be feasible, e.g., through application of the
the G\"artner-Ellis theorem
 \cite{denHollander2000,touchette2009,dembo2010,touchette2011}.

\begin{figure}[htb]
  \includegraphics{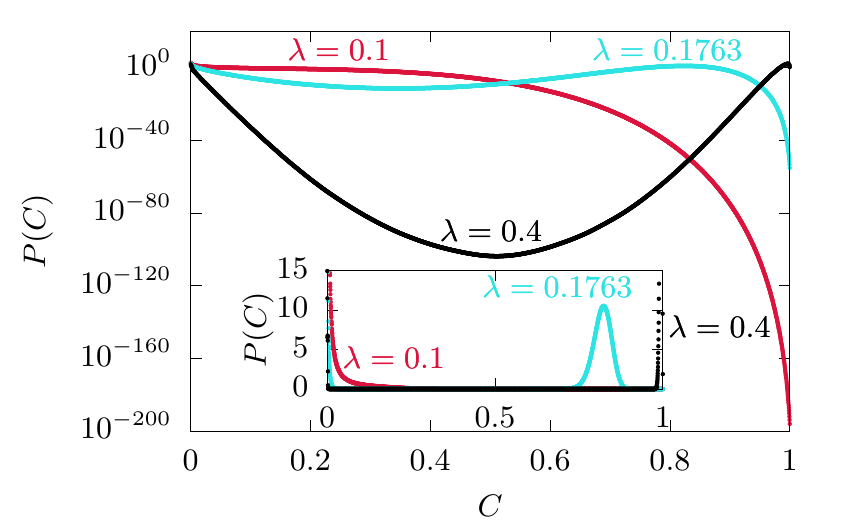}
  \caption{(color online)
Probability density of total infections $C$ for $\mu = 0.14$ and $N=3200$
    for different values of $\lambda$, below, at, and above the critical
    threshold. Linear scale in inset.}
  \label{sw_multi_t}
\end{figure}

So far, we have considered only the critical point
$\lambda\approx \lambda_c$.
A comparison with the other values of the transmission probability is shown in
\autoref{sw_multi_t}.
As one would expect, higher values of $\lambda$ lead to an
increased probability for larger values of $C$.
Correspondingly,
lower values of $C$ become far more
likely for the lower transmission probabilities.
Note that for all three considered values of $\lambda$, we observe
rather high probabilities
for $C\approx 0$, since only one node is infected in the beginning.
But this is only an effect emerging from the initial condition
and can be ignored when discussing the main part of $P(C)$.
Even if one started with a larger number of initially infected nodes,
this will only affect the height of the peak for small values of $C$
and the overall weight of the part for $C\gg 0$, but not the shape.

\subsection{Correlations}

We want to further analyze the properties of typical and atypical
outbreaks to obtain insight into their structure and maybe even
identify possible causes for extreme events.
For this purpose, we store during the entropic sampling
for each WL interval
$200\,000$
time evolutions of outbreaks, at steps evenly spaced out
in the entropic sampling Monte Carlo time.
We store for these time evolutions
the fraction  $i$ of infected, the fraction $s$ of susceptible
and the fraction $c$ of  the
total fraction of so-far infected nodes during the time evolution,
see \autoref{sec-sir-model}.
Below, $T$ will denote either of these quantities
and we call
\begin{equation}
  T = \left( T\left[0\right],\dots,T\left[t_{\max}-1\right]\right)\,.
\end{equation}
a \emph{time series}.

Each time series is binned according to its energy $E$, i.e., $C$ here, or
$M$ in Sec.~\ref{sec:M}. We denote by
$b_E$ the set of a number $B_E$ of time series collected for histogram
bin $E$, i.e.,
\begin{equation}
  b_E = \left\{T_0^{E},\dots,T_{B_E - 1}^{E}\right\}\,.
  \label{bins}
\end{equation}

\begin{figure}[htb]
  \includegraphics{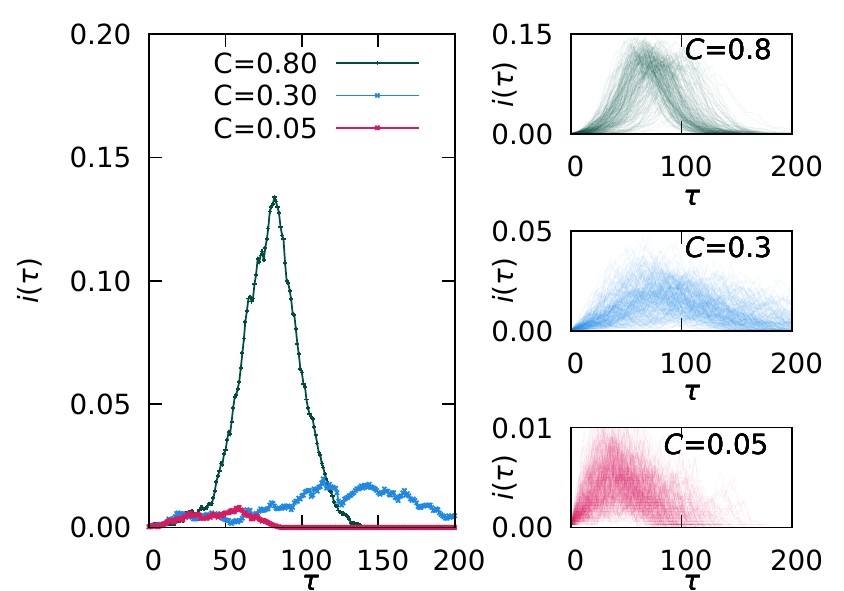}
  \caption{(color online)
Fraction of infected nodes $i$ as function of time $\tau$ for
    three different values of $C$.
    The plot on the left shows a single time series for each $C$ value as
    examples, whereas the plots on the right each show 250 time series
    for their respective $C$ value.}
  \label{example-curves-infected}
\end{figure}

As an example, in \autoref{example-curves-infected} we show a collection
of time
series $i(\tau)$ for three different values of the cumulative fraction $C$
of infected individuals, for  $N=3200$ and $\lambda=\lambda_c$.
One can observe that the
infection can last way longer for medium values of $C$ and dies out quicker
for very large or low values of $C$.

In the next subsection, we
use heat maps to investigate how similar the time series are when
comparing them pair-wise. Afterwards, we
calculate other measurable quantities of the time series
to relate them to the values of $C$ they exhibit, respectively.

\subsubsection{Disparity heat maps \label{ref:disparity}}

To measure how similar the time series are to each other,
we first normalize each series by dividing through its maximal value
encountered during the outbreak.
This way we can better compare the shape of the time series and are not
comparing their magnitudes.

We define a distance $d$ for two normalized  time series
$T, T^\prime$ as
\begin{equation}
  d(T, T^\prime) := t_{\max}^{-1}
  \sum_{\tau = 0}^{t_{\max} - 1}
  \left|   T[\tau]-T^\prime [\tau] \right|\,.
\end{equation}

We define the \emph{disparity}
$V_T(E,E')$ between time series from $E$ and $E'$ as the
averaged distance $d\left(T_\alpha^{E},T_\beta^{E'}\right)$
for pairs of time series taken from the bins
$E$ and
$E^\prime$, respectively.
Here,  we used 500 time
series per bin,
drawn randomly from all saved time series that were collected for the
respective bin.
Hence a total of up to $1\,600\,000$ time series are used in each analysis.

\begin{figure}[htb]
  \includegraphics{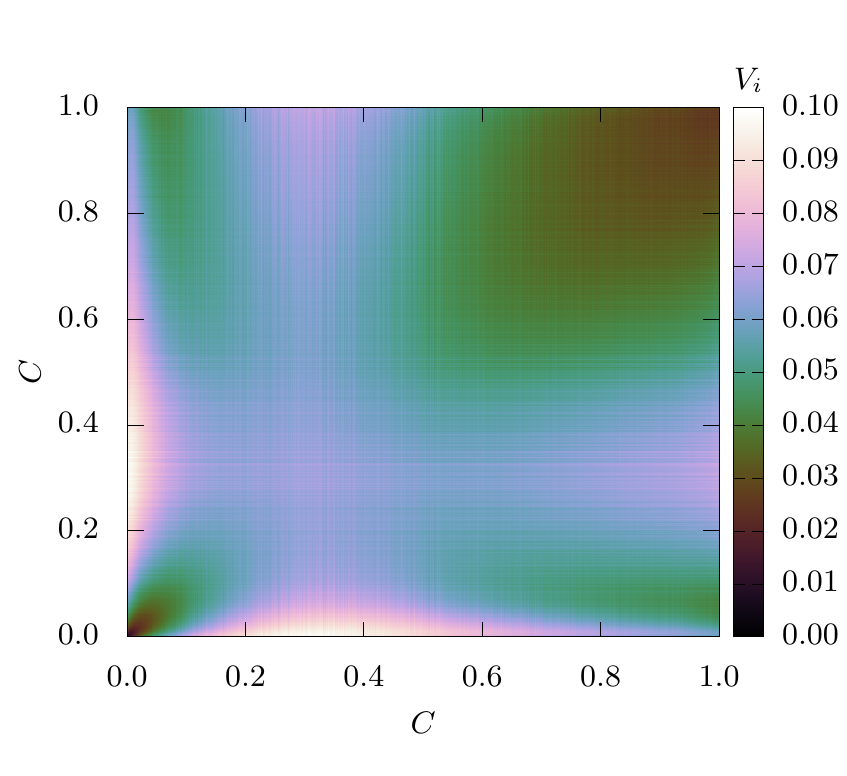}
  \caption{(color online)
Disparity $V_i$ of the time series $i(\tau)$ of the fraction
    of infected individuals for pairs of time series binned with respect to
    their total
    fraction $C$ of infections,
    for $N=3200$, $\mu=0.14$ and $\lambda=0.1763$ }
  \label{V_abs_i}
\end{figure}

In \autoref{V_abs_i} we show the disparity  $V_{i}$, for the
fraction of infected, for $\lambda\approx\lambda_c$ color-coded, i.e.,
in form of a heat map. Note that we are able to show the disparities
over the full range of possible values for $C$, which is only possible
because we applied the large-deviation approach. When using simple
sampling instead, only a very small range of values near $C\approx 0$ and
near $C\approx 0.8$ would be accessible.
Here we are able instead to identify three different regions.
The first is located in the range
$0 \leq C \leq 0.1$, the second is $0.1 < C \leq 0.5$ and the third
for $C > 0.5$.
Region  one and three seem to consist of time series where the shapes
are similar to each other, within the region, visible by the dark
color around the diagonal in the heat map. But they are quite different
to other regions.
Region two seems to consist of different time series that are not even
that similar to time series from its own region, i.e., here we observe
strong fluctuations from time series to time series.
Note that in \autoref{example-curves-infected} the time series are
selected from these regions, and
thus illustrate their behaviors.
\autoref{example-curves-infected} tells us that
time series from the first region exhibit only small
fractions of infected and the outbreak dies out quickly. For the second
region, we observe medium strong outbreaks, but they may take very
long and the shapes and durations fluctuate strongly. In the third region,
many individuals get infected during the outbreak,
leading to an even larger fraction
of individuals infected at the same time, but the outbreak finished
more quickly than in region two.

For $\lambda=0.1$ the heatmap looks similar (not shown), though now region 2 is shifted
towards larger values of $C$, i.e., the region is $0.15 < C \le 0.8$,
and the other regions change accordingly.

For $\lambda = 0.4$ the heatmap also looks similar (not shown). Here
the second region is much smaller, i.e. $0.075 < C \le 0.21$,
and the other regions change accordingly.

\begin{figure}[htb]
  \includegraphics{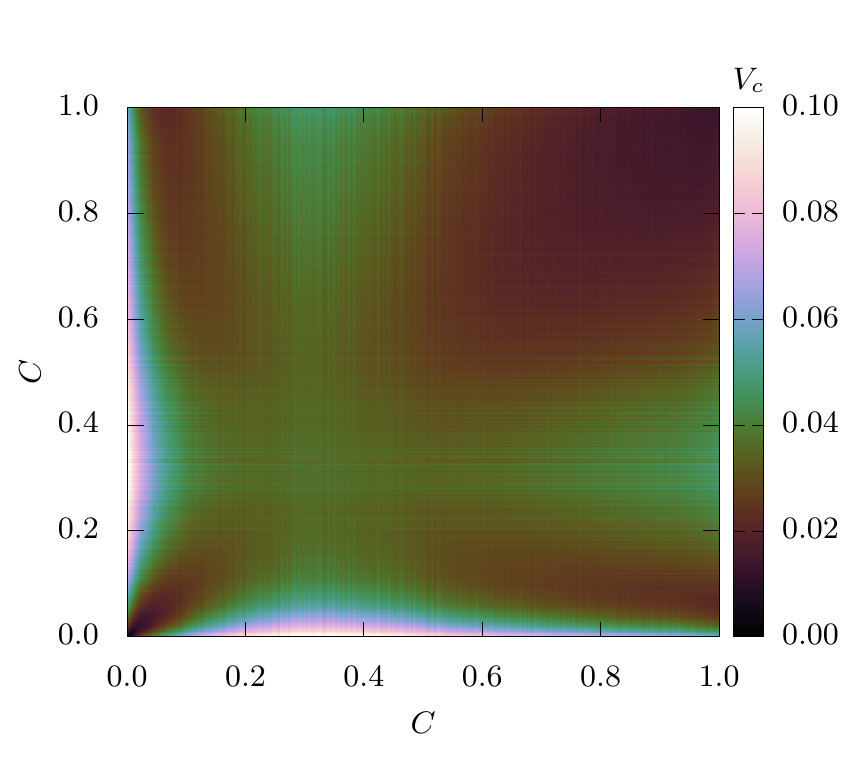}
  \caption{(color online)
Disparity $V_c$ of the time series $c(\tau)$ of the fraction of cumulative infections for pairs of time series binned with respect to their total fraction
    $C$ of infections for $N=3200$, $\mu=0.14$ and $\lambda=0.1763$ } 
  \label{V_abs_ever}
\end{figure}

In \autoref{V_abs_ever} we show the disparity heat map for the
time series of cumulative infections, i.e., $V_c$.
This heat map adumbrates the three regions as well, though they are much less pronounced. Thus, to compare the dynamics of infections, the
current fraction of infections allows for a better insight compared to
the cumulative fraction of infections.
For $\lambda=0.4$ we see the same, though the region borders changed, as mentioned
previously. For $\lambda=0.1$ the heat map also hints at the regions previously
mentioned, however region 2 is very faint and barely visible.

\subsubsection{Conditional density\label{ref:conditional}}

In order to study the relation of other measurable properties $Q$
of the time series to
their values of $C$, we study conditional densities.
Again we bin each time series according to its energy $E$,
here $E=C$, (see \autoref{bins})
and then obtain a normalized histogram $\rho_T(Q|E)$
of $Q$ given $E$.
Again  $T$ will be either $i$, $s$, $c$, or omitted, if suitable.

For the measurable quantities $Q$ we considered
\begin{itemize}
\item $M$ as defined above, i.e.,
  the maximum of the fraction $i$
  of currently infected nodes during an outbreak;
\item the time steps $\tau_{\max}$ until
  the maximal value of the fraction $i$ of infected
  is reached.  This measures the time scale it takes
  for an outbreak to reach its maximum activity.
  This is interesting for practical purposes, as it translates
  to the time where
  the maximal healthcare capacity is required;

\item
the time steps $\tau_{\min}$
until the minimal value of the fraction $s$ of susceptible nodes is reached.
This means that, after this time, no additional nodes obtained an infection, 
although the recovery of the remaining infected nodes still takes some time.
This quantity is a  measure for the outbreak duration;
\item
the number of time steps $\tau_{10}^{90}$ it took
such that the fractions $i$ or $c$ here,
raised from 10\% to 90\% of its maximal value, respectively.
In the few cases where, for analyzing $i$, this occurred
several times, we only consider the duration of the first occurrence.
These time scales quantify how long the outbreak is very active;
\item the fraction $f_{\rm sw}$ of how many neighbors of an infected
  node are infected through a long-range edge, i.e., along those
  edges which were rewired during graph generation. Hence
  \begin{equation}
  f_{\rm sw} = \frac 1 N \sum_{i,I_i>0} \frac{ L_i}{I_i}\,,
\end{equation}
  where $I_i$ is the number of neighbors of node $i$ infected by node $i$
  and $L_i$ the number of neighbors of node $i$ infected through long-ranging
  edges;
\end{itemize}

\begin{figure}[htb]
  \includegraphics{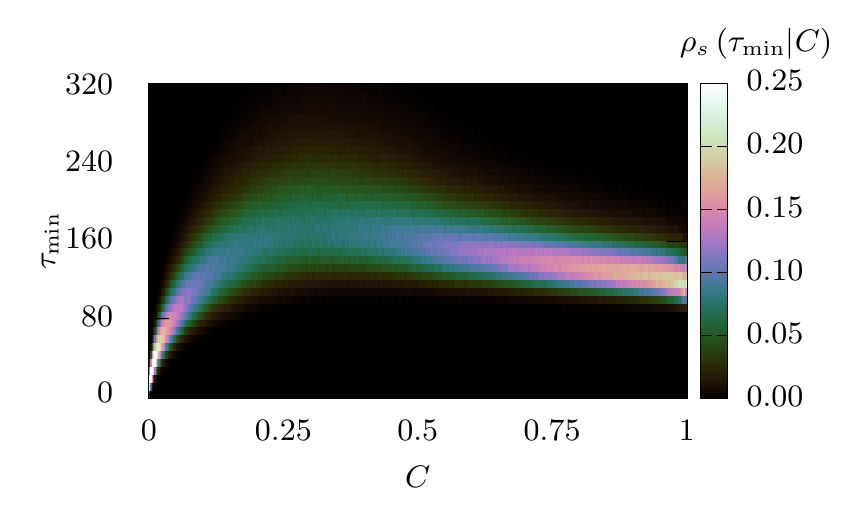} 
  \caption{(color online)
Conditional density $\rho_s\left(\tau_{\min}|C\right)$,
    which shows the probability
    of $\tau_{\min}$, i.e., how many time steps it takes until the last
    node got infected during an outbreak,
  for any given $C$. The system size is $N=3200$, the recover probability $\mu=0.14$ and the transmission probability $\lambda=0.1763$.
  }
    \label{H2_min_s}
\end{figure}

In \autoref{H2_min_s} we show the distribution $p_s(\tau_{\min}|E)$
for the time
steps $\tau_{\min}$
it took until the minimum of susceptible nodes was reached,
conditioned to the value of $C$.
As one would expect, this distribution is centered at small
times for low values of $C$.
For $0.25 \leq C \leq 0.5$ the disease survives the longest and
exhibits the largest spread in times scales. When
increasing $C$ further, the life time
of the disease decreases again. These results support the insight gained
for the different regions from looking at the sample time series
in \autoref{example-curves-infected}.

For $\lambda = 0.1$ (not shown) the shape is similar, though flatter and
the range where the disease survives the longest stretches  now from
$0.25\leq C \leq 0.75$. Also the disease generally survives longer,
because the pool of
susceptible nodes decreases more slowly.
For $\lambda=0.4$ (not shown) the shape looks even more similar to
the one shown in \autoref{H2_min_s},
though the maximum is now more pronounced and around $C\approx 0.13$
and the disease dies out even  faster, because it rushes more quickly
through the population.

\begin{figure}[htb]
  \includegraphics{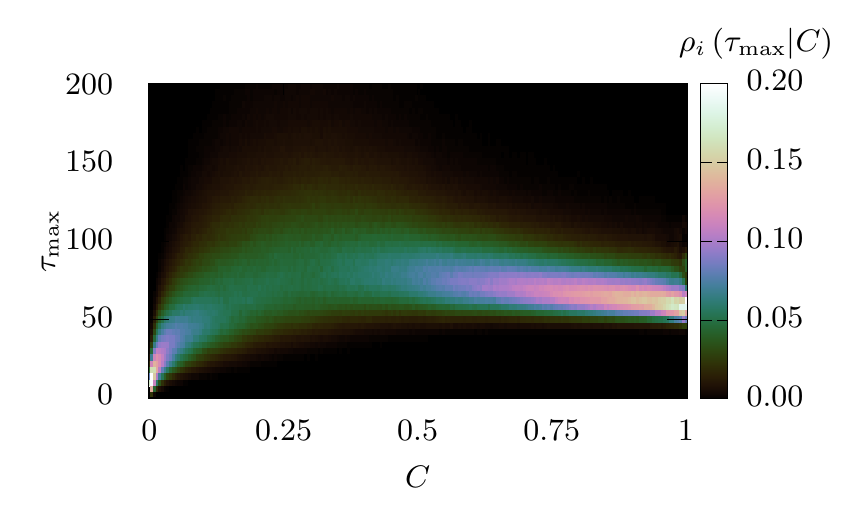} 
  \caption{(color online) Conditional 
density $\rho_i\left(\tau_{\max}|C\right)$,
  which shows the probability of $\tau_{\max}$, i.e., how many time steps it takes
  to reach the peak of the infection time series $i(\tau)$, for any given value of $C$.
  The system size is $N=3200$, the recover probability $\mu=0.14$ and the transmission probability $\lambda=0.1763$
  }
    \label{H2_max_i}
\end{figure}

In \autoref{H2_max_i} the conditional distribution $p_i(\tau_{\max}|E)$
for the time of the peak infection is shown.
As one can see, the shape is similar to the one from \autoref{H2_min_s}.
From this similarity one can conclude $\tau_{\min}\sim\tau_{\max}$, i.e.,
the longer the disease lasts, the later the peak of the
fraction of infected occurs.
We find the same results for $\lambda = 0.1$ and $\lambda=0.4$
(not shown).

Looking back at the diagonals of \autoref{V_abs_i} and \autoref{V_abs_ever}
we notice, that we observe the largest value of $\tau_{\max}$ as well as the 
largest variance of $\tau_{\max}$ for the range where 
the \emph{disparity} $V$ 
was the highest, i.e., where the respective time series exhibited the most 
variation. 
This makes sense intuitively and reminds one of the behavior of
systems near critical slowing down.   

\begin{figure}[htb]
  \includegraphics{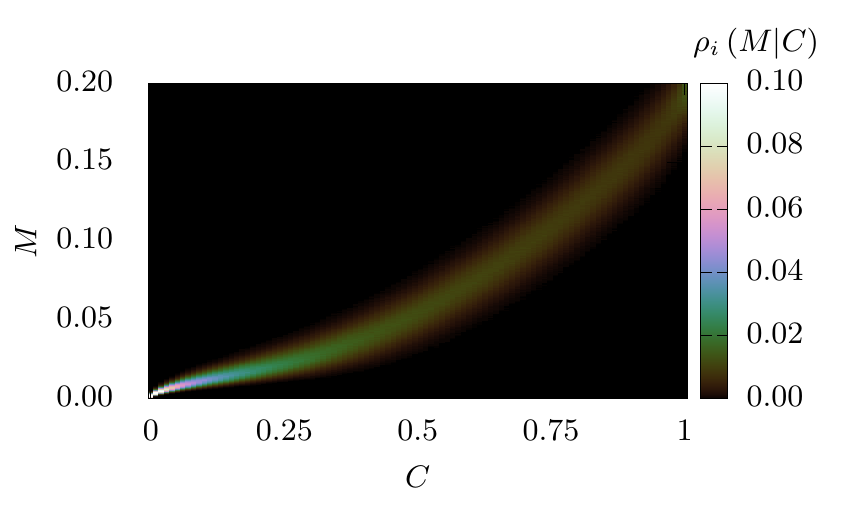} 
  \caption{(color online)
Conditional density $\rho_i\left(M|C\right)$, which shows the probability of $M$, i.e., the
  maximum of the time series $i(\tau)$, for any given value of $C$.
  The system size is $N=3200$, the recover probability $\mu=0.14$ and the transmission probability $\lambda=0.1763$.}
    \label{H2_max_v}
\end{figure}

In \autoref{H2_max_v} the conditional distribution $p_i(M|C)$
for the maximum of nodes, which are in the infected state at the same time,
is shown. Here, the three regions, which where visible in
Figs.~\ref{H2_min_s} and \ref{H2_max_i} are not apparent. Instead
one observes a generally monotonous relation between $C$ and the center of
the distribution of $M$. Still, this is compatible with
the outbreak examples shown in \autoref{example-curves-infected}.
The same behavior can be observed for $\lambda=0.1$ (not shown),
though the peak value $M$ is
generally lower and for $\lambda=0.4$ (not shown) the peak value
is higher.

\begin{figure}[htb]
  \includegraphics{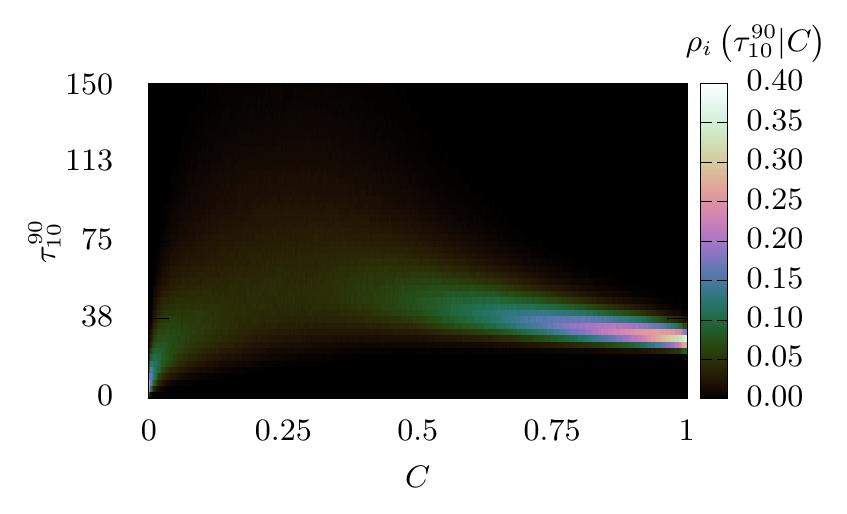}
  \caption{(color online)
Conditional density $\rho_i\left(\tau_{10}^{90}|C\right)$,
  which shows the probability of $\tau_{10}^{90}$, i.e.,
  the duration between reaching 10\% and 90\% of the maximum of $i(\tau)$,
  for any given value of $C$.
  The system size is $N=3200$, the recover probability $\mu=0.14$ and the transmission probability $\lambda=0.1763$. }
  \label{10_to_90_infected}
\end{figure}

In \autoref{10_to_90_infected} we show the conditional density
$\rho_i\left(\tau_{10}^{90}|C\right)$ for the duration of the
most-active phase of the outbreak.
As one can see, the spread of the duration times
is the largest for region two, where
the time series also looked more chaotic
(see \autoref{example-curves-infected} for $T=i$).
If we look at the $\tau^{90}_{10}$ for $T=c$ (not shown)
the plot looks quite similar, though the durations are
a bit longer overall.
For $\lambda=0.1$ the heat map looks similar, but more like a half moon,
and the values scatter most around $C\approx0.5$. Also the values
are a bit higher overall.
For $\lambda=0.4$ the basic shape also looks similar, but the slope is steeper
at the beginning. It also flattens out sharply at about $C\approx 0.5$, i.e., at a
relatively small value. The values are lower overall compared to the other two
values of $\lambda$.

We are also interested in the effect of the long ranging connections.
For this we measure the fraction of long ranging edges that
caused an infection.

\begin{figure}[htb]
  \includegraphics{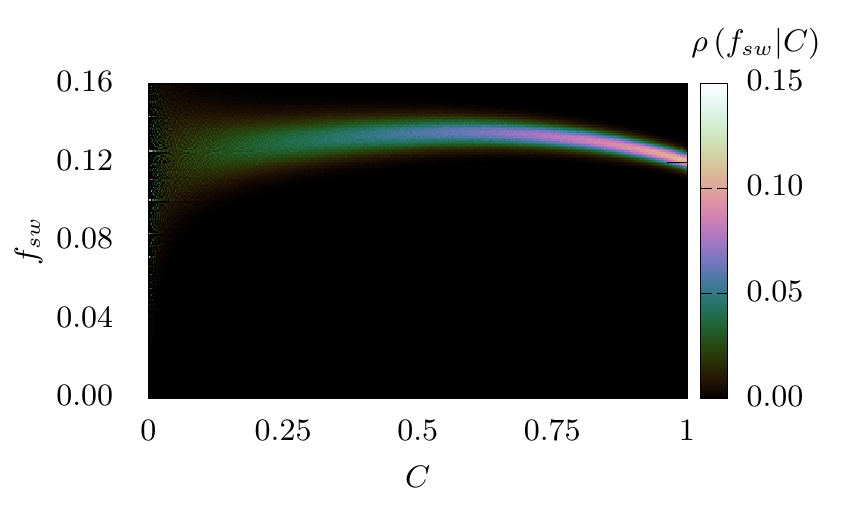}
  \caption{(color online)
Conditional density $\rho\left(f_{\rm sw}|C\right)$, which shows the probability
  of $f_{\rm sw}$ for any given $C$.
  The system size is $N=3200$, the recover probability $\mu=0.14$ and the transmission probability $\lambda=0.1763$}
  \label{ever_infected_sw_frac}
\end{figure}

In \autoref{ever_infected_sw_frac}, we show the conditional distribution
$\rho\left(f_{\rm sw}|C\right)$ of the fraction of infections through
long-range edges, i.e., those edges which are responsible for the
small-world behavior.
For small values of $C$, the values of $f_{\rm sw}$ scatter strongly,
because here the disease dies
out very quickly and thus $f_{\rm sw}$ is obtained by averaging over only very few contagions.
Overall we see a weak correlation of $C$ and $f_{\rm sw}$,
where $f_{\rm sw}$ increases slightly for
larger $C$ until $C\approx 0.75$. Thus, we see a weak effect that in order
to see a global pandemic, the spread has to go to a slightly larger extend
through long-range
connections and spreads a little bit less locally.
This, even within such a simple model,
supports the often used real-word strategy to suppress with higher
priority long-distance traveling. Interestingly, because this correlation is
seen also for rather large values of $C$, this would also help at least
a bit even if a pandemic has broken out already, not only in an early stage
to prevent pandemic outbreaks.
On the other hand,
to actually reach all individuals, i.e., for even larger values of $C$,
 many local edges have to be involved since there are still
many nodes which only have local neighbors. This explains the decrease
visible for large values of $C$.
We first suspected that the overall weak correlation was due to our high
probability $p=0.1$ of rewiring, because of
which one third of the nodes are adjacent to at least one
long-range edge.

\begin{figure}[htb]
  \includegraphics{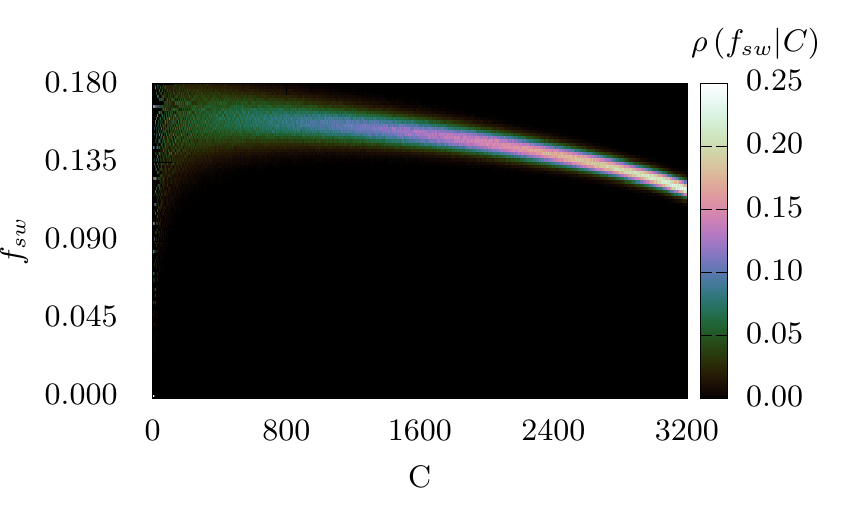}
  \caption{(color online) Conditional density 
$\rho\left(f_{\rm sw}|C\right)$, which shows the probability
  of $f_{\rm sw}$ for any given $C$.
  The system size is $N=3200$, the recover probability $\mu=0.14$ and the transmission probability $\lambda=0.1$}
  \label{ever_infected_sw_frac_0.1}
\end{figure}

\begin{figure}[htb]
  \includegraphics{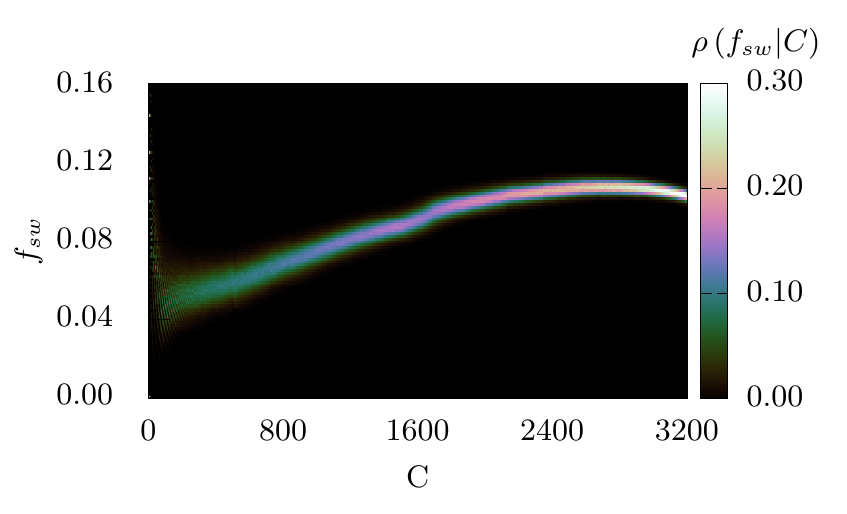}
  \caption{(color online) Conditional density 
$\rho\left(f_{\rm sw}|C\right)$, which shows the probability
  of $f_{\rm sw}$ for any given $C$.
  The system size is $N=3200$, the recover probability $\mu=0.14$ and the transmission probability $\lambda=0.4$}
  \label{ever_infected_sw_frac_0.4}
\end{figure}

The plots for $\lambda=0.1$ \autoref{ever_infected_sw_frac_0.1} and
$\lambda=0.4$ \autoref{ever_infected_sw_frac_0.4}, however, paint a different picture.
For the transmission probability ($\lambda=0.1$) below the critical value, we observe an anti correlation
between $f_{\rm sw}$ and $C$, whereas we see a correlation for the transmission probability ($\lambda=0.4$) above
the critical value.
The latter is what we expected, since the transmission via long ranging edges infects more distant nodes,
which can start new infection clusters.

\section{Maximum fraction  $M$ of currently infected\label{sec:M}}

Next, we study the large-deviation properties with respect to
the maximum fraction $M$ of simultaneously infected nodes, in
a similar way as we have done for $C$.
Although we have seen a strong relationship between $C$ and $M$
in \autoref{H2_max_v}, we will show below that
not all results obtained for $E=C$
transfer directly to the case $E=M$.
Note that for obtaining these results
we had to perform completely independent large-scale
simulations with energy $E=M$
in order the access also the outbreaks which have
a rare behavior with respect to $M$.
For our largest system size we evaluated $M$ about $1.5 \times 10^{10}$
times during entropic sampling and  Wang Landau combined,
which is thus the total number of local MC attempts.

The parameters we use for the simulations are presented
in Tab.~\ref{tab:paramM}, for the
different networks sizes $N$ and values of the transmission probability $\lambda$.

\begin{table}[h!]
  \begin{tabular}{|ll|lll|}
  $\lambda$ & N    & approach  & $\#I$ & 	$B$	 \\ \hline
  0.1763    & 400  & WL                       & 8     &	66	\\
  0.1763    & 800  & WL                       & 20     &	166	\\
  0.1763    & 1600 & REWL                     & 24     &	512	\\
  0.1763    & 2400 & WL                       & 24     &	1024	\\
  0.1763    & 3200 & WL                       & 24     &	2048\\
  0.1763    & 6400 & WL                       & 42     &	3072\\
  0.1763    & 6400 & REWL                     & 6      &	4096\\
  0.1    & 1600  & REWL                       & 24     &	512	\\
  0.4    & 1600  & REWL                       & 24     &	512	\\
  \end{tabular}
  \caption{\label{tab:paramM}
    Parameters for the simulations: transmission probability $\lambda$,
    number of nodes $N$, the approach used, the number $\#I$ of intervals used
    in the WL or REWL sampling  and the number $B$ of exchanges
    performed per MCMC attempt for the
    arrays $\xi_\lambda$ and $\xi_{\mu}$ of random numbers.}
  \end{table}

For the noncritical values of $\lambda$, namely
 $\lambda=0.1$ and $\lambda=0.4$,  we perform
simulations only for a system size of $N=1600$. We originally wanted to use 
a system size of $N=3200$ here, though we encountered problems during 
the sampling  of $\lambda=0.4$.
Note that we sample $N=6400$ ($\lambda=\lambda_c$) 
with a combination of WL and REWL.

\begin{figure}[htb]
    \includegraphics{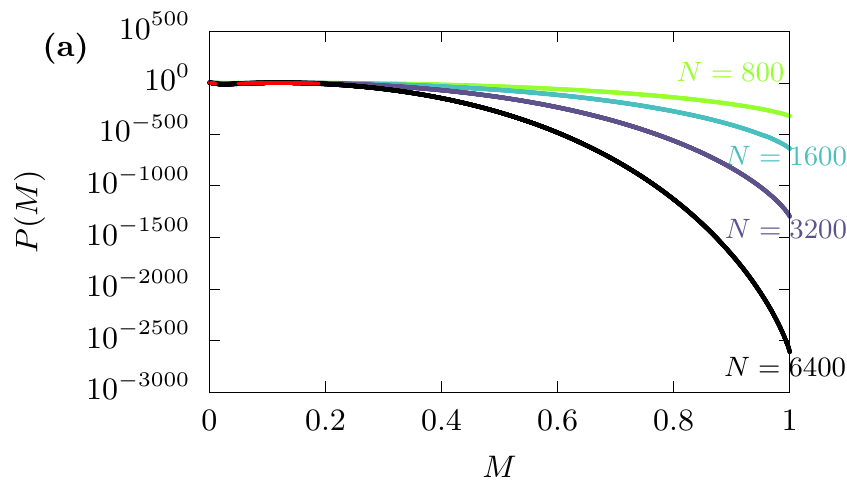}\\
    \includegraphics[width=0.49\linewidth]{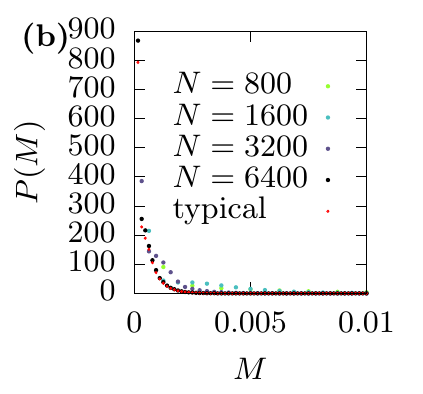}
    \includegraphics[width=0.49\linewidth]{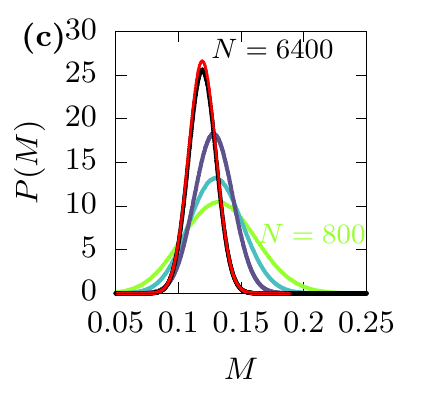}
    \caption{(color online) Probability density of maximum fraction $M$
  of simultaneously infected
  for $\mu = 0.14$ and $\lambda=0.1763$. (a) shows the full distribution
  in logarithmic scale, while (b) and (c) highlight the two peak regions
  in linear scale.  For (a) and (c) the largest and smallest values of 
  $N$ are labeled,  in between the values behave monotonically,
  except for a small area around $M=0.125$, where the order is  reversed, see (c). 
  The used values were $N\in\{800,1600,3200,6400\}$.
  We also included the typical-event sampling results for $N=6400$ in red.}
    \label{sw_multi_N_other}
  \end{figure}

In \autoref{sw_multi_N_other} the probability density $P(M)$
is shown for different system sizes $N$ and $\lambda=\lambda_c$.
Here, we never encountered an outbreak that lasted longer than
our simulation time $t_{\max}$. This means, we have chosen $t_{\max}$
large enough to understand the behavior of $M$ over the full
range of possible values sufficiently.
Using the large deviation approach we are able to measure
probabilities ranging over 2500
decades.
Thus, if one were to sample the same distribution using only a 
typical-event sampling
approach, one would need at least $10^{2490}$ times as much
computational power as we used, which is clearly infeasible.

We also include the typical-event sampling results for $N=6400$ where we use
$10^{10}$ samples. Looking at the logarithmic probabilities
the large deviation results and typical-event sampling agree very well.
However, at the peaks in the linear scale, seem to be discrepancies.
These come from errors that accumulate during the glueing of the 
large deviation intervals, which come inevitably into play
because of the sampling fluctuations. But the deviations are
small: the first peak would align perfectly if we add 0.04 to our 
typical-event sampling data in the logarithmic range, while the 
second peak would align perfectly if we  subtract 0.015 from the 
typical-event sampling data in the logarithmic range. 
Looking at the number of magnitudes this measurements covers in the 
probability density range, these errors are really rather small, though 
they become noticable in the linear range.

The probability density function exhibits a peak at $M=\frac{1}{N}\approx 0$
for the same reasons as there is a peak at $C\approx 0$ in $P(C)$.
We also observe a peak at about $M=0.125$.
This means, if the
infection survives the first few steps one can assume that
typically about 12.5\% of the network will be infected at
the same time at some point  of the outbreak. This determines
the capacity of the health care system required to cope with typical outbreaks.
If one wants to be prepared for large atypical outbreaks, the
tails become important. Nevertheless,
for substantially larger values of $M$ the probability density becomes
very small. Unsurprisingly,
the least likely case is that the entire network is infected at the same
time at some point of the outbreak. Clearly, these extreme tails of the
distribution are not relevant for practical applications, but from
a fundamental and scientific viewpoint, it is pleasing to be able
to calculate the distribution over its full support.

\begin{figure}[htb]
  \includegraphics{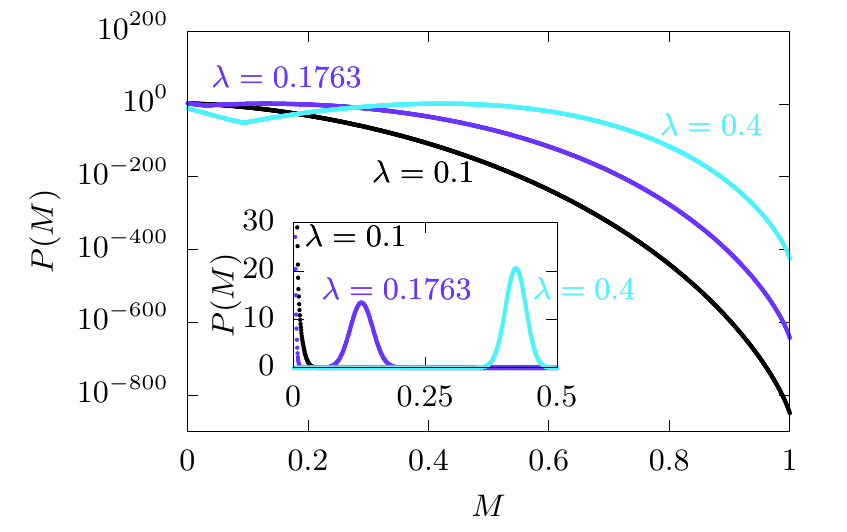}
  \caption{(color online)
Probability density of the maximum fraction $M$ of simultaneously infected for $\mu = 0.14$ and $N=1600$ and different $\lambda$. Linear scale in inset}
  \label{max_infectedsw_multi_t}
\end{figure}

In \autoref{max_infectedsw_multi_t} the pdf is shown for different
transmission probabilities $\lambda$.
Note the kink for $\lambda=0.4$. We are only able
to resolve this interesting point where the distribution
seems to be not differentiable by using the replica exchange Wang-Landau
algorithm here.
Unsurprisingly, larger transmission probabilities $\lambda$
lead to an increase of the probability to observe larger values of $M$.

\begin{figure}[htb]
  \includegraphics{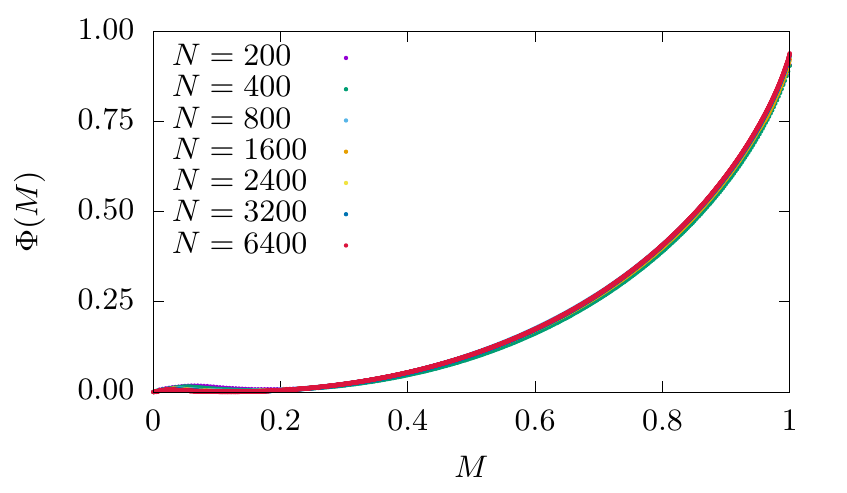}
  \caption{(color online)
Empirical rate function $\Phi(M)$ for multiple system sizes $N$ and $\mu=0.14$ and $\lambda = 0.1763$}
  \label{rate_function_max_infected}
\end{figure}

In \autoref{rate_function_max_infected} we show the rate function as measured for different system sizes.
Clearly, the rate functions all agree very well. Basically, no
finite-size effects are visible in contrast to the case of $P(C)$.
This means one can use the rate function to predict the pdf for any
system size $N$.
We verified that by using the rate function calculated for $N=3200$ to
accurately predict the pdf for $N=400$ and $N=2400$.
Thus, it is not necessary to perform
any extrapolation of the rate function. This means
that for $P(M)$ our numerical results also indicate
that the large-deviation principle is fulfilled. Nevertheless,
 the kink
visible for $\lambda=0.4$ hints that in the pandemic phase for larger
values of $\lambda$, the mathematical properties of the rate function
could pose some problems to an analytic treatment.

\subsection{Correlations}
\subsubsection{Disparity heat maps}

\begin{figure}[htb]
  \includegraphics{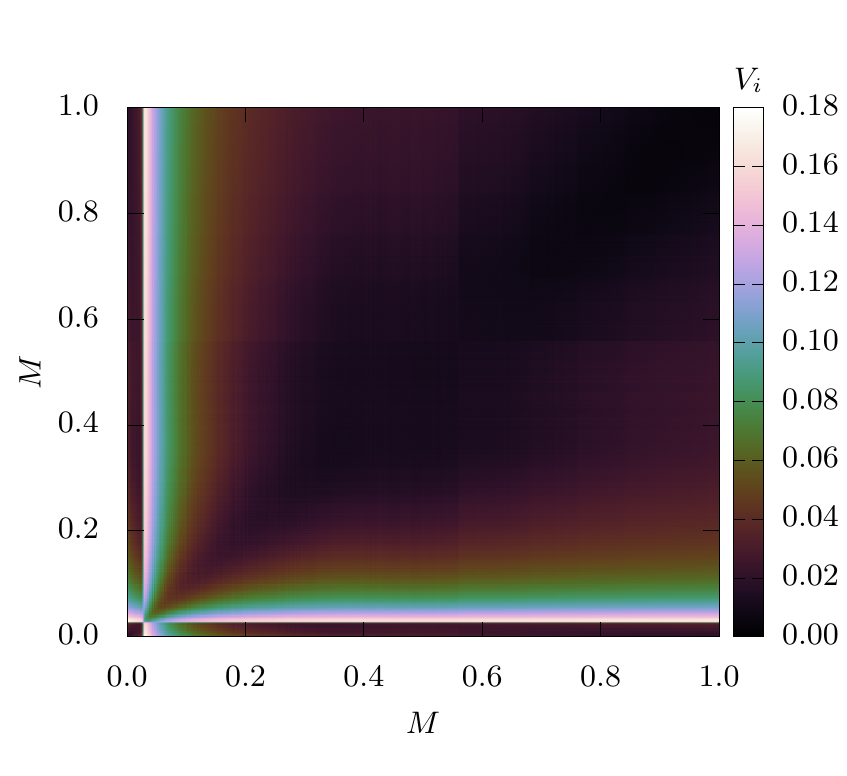}
  \caption{(color online)
Disparity $V_i$ of the time series $i(\tau)$ of the fraction of infected individuals for pairs of time series binned with respect to their total fraction $C$ of infections for $N=3200$, $\mu=0.14$ and $\lambda=0.1763$}
  \label{max_infected_V_abs_i}
\end{figure}

Again we study the disparity of outbreaks, now for pairs of
times series of
outbreaks classified according their values of $M$, respectively.
In \autoref{max_infected_V_abs_i} the disparity  $V_i$
for the time series of the fraction of infected is shown, see
\autoref{ref:disparity}.
As in \autoref{V_abs_i}, we again see the three regions,
first for very small values $0 \leq M < 0.035$, the second for
$ 0.035 \leq M < 0.08$ and the third for $M\geq 0.08$. Hence,
in contrast to the case of the cumulative fraction $C$ of infections,
two of the three regions are visible on a much smaller range of values.
For $\lambda=0.1$ the heat map (not shown) looks quite similar, 
though the second region is shifted towards even lower $M$.
Note that for $\lambda=0.4$ we are only able to ensure convergence 
of the distribution for $N=1600$ and  therefore are 
comparing different system sizes. Still, the differences of the disparity
plots between such rather
large system sizes are very small, as we observe in general during our study.
The heat map (not shown) also looks rather similar, 
though the second region is shifted
towards larger $M$.

\begin{figure}[htb]
  \includegraphics{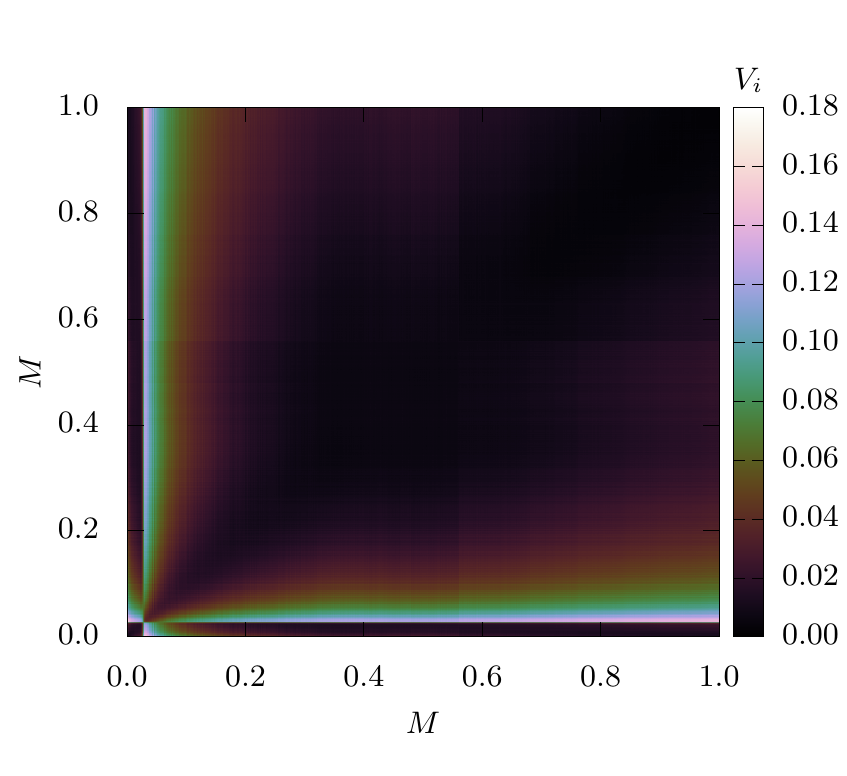}
  \caption{(color online)
Disparity $V_c$ of the time series $c(\tau)$ of the fraction of cumulative infections for pairs of time series binned with respect to their total fraction $C$ of infections for $N=3200$, $\mu=0.14$, $\lambda=0.1763$}
  \label{max_infected_V_abs_ie}
\end{figure}

\autoref{max_infected_V_abs_ie} shows the disparity
when comparing the time series  of the cumulative infections.
It looks quite similar to \autoref{max_infected_V_abs_i}. This is
also in contrast to the case of classifying the outbreaks according $C$,
where the two disparity heat maps $V_i$ and $V_c$ appeared more different.
For $\lambda \in \{0.1, 0.4\}$ the heat maps $V_i$ and $V_c$ (not shown) also look alike.

\subsubsection{Conditional density}

\begin{figure}[htb]
  \includegraphics{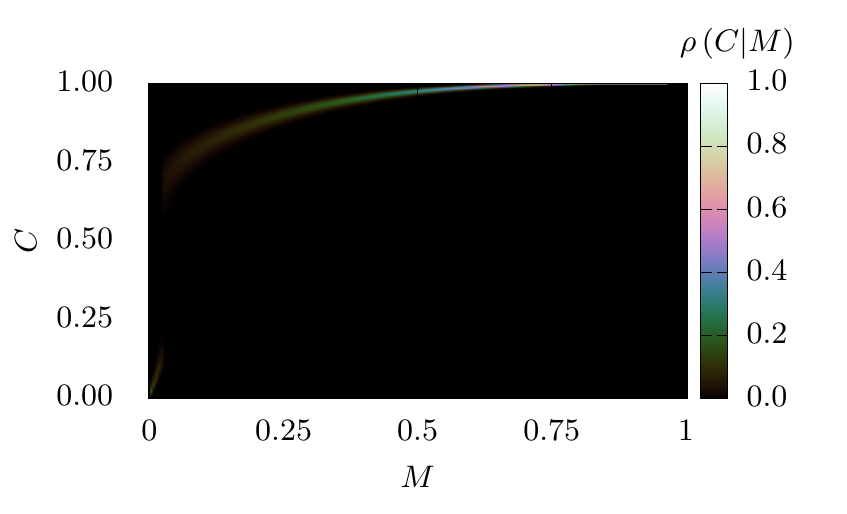}
  \caption{(color online)
Conditional density $\rho\left(C|M\right)$, which shows the probability of $C$, i.e.,
  the total fraction of infections, for any given $M$.
  The system size is $N=3200$, the recover probability $\mu=0.14$ and the transmission probability $\lambda=0.1763$}
    \label{Correlation_M_C}
\end{figure}

Although we have already studied $\rho(M|C)$, we show
in \autoref{Correlation_M_C} the distribution $\rho(C|M)$ of
$C$ conditioned to the
value of $M$. See \autoref{ref:conditional} for details.
Note that the non-zero values  must be located above the diagonal, because for
every outbreak $C\geq M$ holds by definition.

In general, $C$ and $M$ still are  monotonously
related. But one can see, there is a sharp increase of $C$ right
between regions two and three, near $M\approx 0.08$.
In this small interval, outbreaks with relatively small and 
relatively large values of $C$ lead to the same observed maxima $M$.
Thus, the change from region two to three coincides with a strong change
and large fluctuations, similar to the behavior of physical phase
transitions.
This difference
shows that when biasing with respect to $M$ one analyses the behavior
in a different way than when biasing with respect to $C$ as in the
previous section. The reason is that in fact there is an underlying
joint distribution $P(C,M)$, but for sampling this one even down
to the tails, one would have to apply a kind of two-dimensional
rare-event sampling approach which is currently out of reach for the present
problem and the considered graph sizes.

The observed behavior is even more pronounced for $\lambda = 0.4$
(not shown).
Here this behavior also corresponds to the position of the kink in the pdfs
from \autoref{max_infectedsw_multi_t}.
With respect to our algorithmic approaches,
we believe that these large fluctuations are
a reason why our initially applied standard Wang-Landau
approach
did not converge and we had to use the replica exchange algorithm.

On the other hand, for $\lambda=0.1$ we do not see such a jump
and strong fluctuations in the
conditional density (plot not shown). Thus, in the non-pandemic phase,
the behavior seems to be simpler, even when including the
large-deviation behavior in the analysis.

\begin{figure}[htb]
  \includegraphics{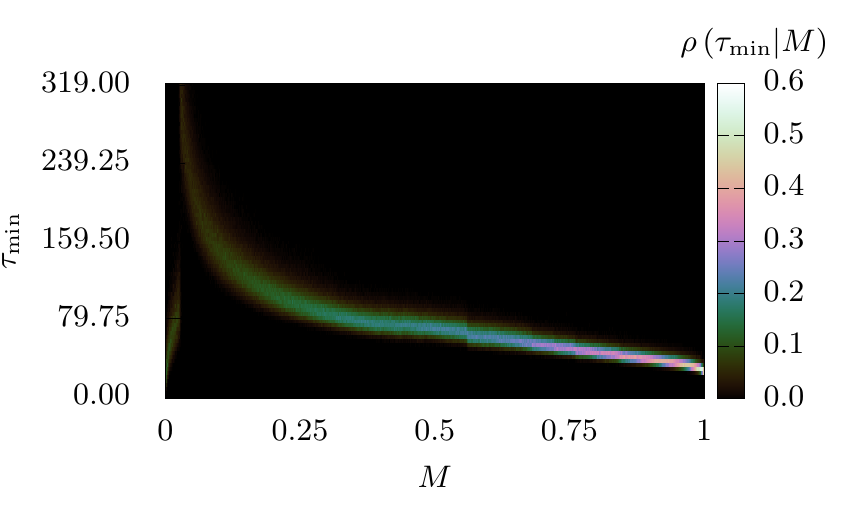}
  \caption{(color online)
Conditional density $\rho_s\left(\tau_{\min}|M\right)$, which shows the
  probability of $\tau_{\min}$, i.e., how many time steps it takes until the last node is infected
  during an outbreak.
  The system size is $N=3200$, the recover probability $\mu=0.14$ and the transmission probability $\lambda=0.1763$}
    \label{H2_max_infected_min_s}
\end{figure}

In \autoref{H2_max_infected_min_s} we show for $\lambda=\lambda_c$
the conditional
distribution
$\rho_s(\tau_{\min}|M)$ of the time scale $\tau_{\min}$ it takes until
the outbreak stops to grow, see \autoref{ref:conditional}.
Here we see a sharp peak at the position $M$, where the jump occured for  \autoref{Correlation_M_C}.
This is consistent with the above observations, where region two was also
associated with the longest outbreak durations. Also, reaching
heavy outbreaks in terms of the fraction $M$ of infections occurring at
the same time, does not at all coincide with long-lasting  outbreaks,
since most of the figure exhibits a negative correlation.
For $\lambda=0.4$ the result (not shown) looks quite similar, though
the outbreak dies down quicker overall and the peak is shifted
a small bit to larger values of $M$.
For $\lambda=0.1$ the plot (not shown) looks quit similar as well,
though the peak is shifted slightly towards smaller values of $M$.

\begin{figure}[htb]
  \includegraphics{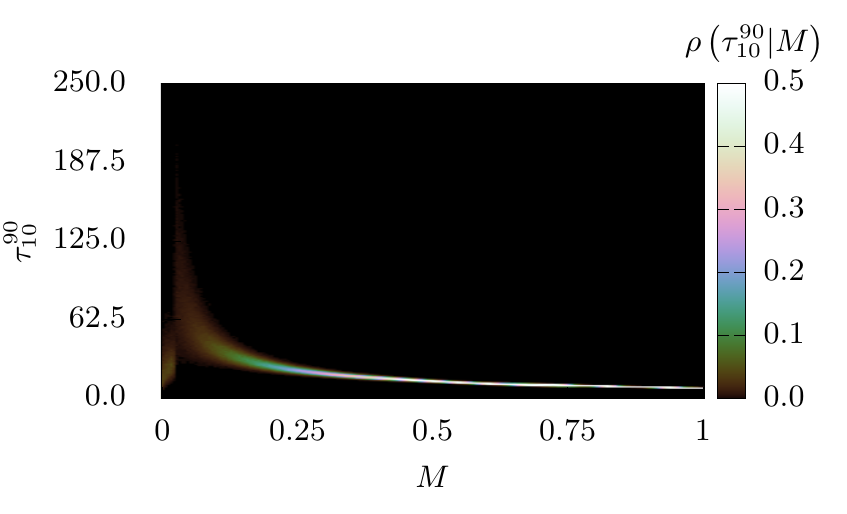}
  \caption{(color online)
Conditional density $\rho_i\left(\tau_{10}^{90}|M\right)$,
  which shows the probability of $\tau_{10}^{90}$, i.e., the duration
  between reaching 10\% and 90\% of the maximum of $i(\tau)$, for any given value of $M$.
  The system size is $N=3200$, the recover probability $\mu=0.14$ and the transmission probability $\lambda=0.1763$}
  \label{max_infected_10_to_90_infected}
\end{figure}

In \autoref{max_infected_10_to_90_infected} we show the conditional density $\rho_i\left(\tau_{10}^{90}|M\right)$ for the duration $\tau_{10}^{90}$
of the highest-activity outbreak phase, see \autoref{ref:conditional}.
We can also see a peak which corresponds
to region two.  Beyond the peak, $\tau_{10}^{90}$ is negatively
correlated, which appears meaningful, since the larger the peak
of the infections in the epidemic phase,
the less time it takes for the outbreak to evolve.
For $\lambda=0.4$ the plot (not shown) looks quite similar, though,
as one would expect,  the durations are shorter over all.
Also the peak is shifted towards larger $M$ again.
For $\lambda = 0.1$ the plot (not shown) looks almost identical
to the plot for $\lambda = 0.1763$.

\begin{figure}[htb]
  \includegraphics{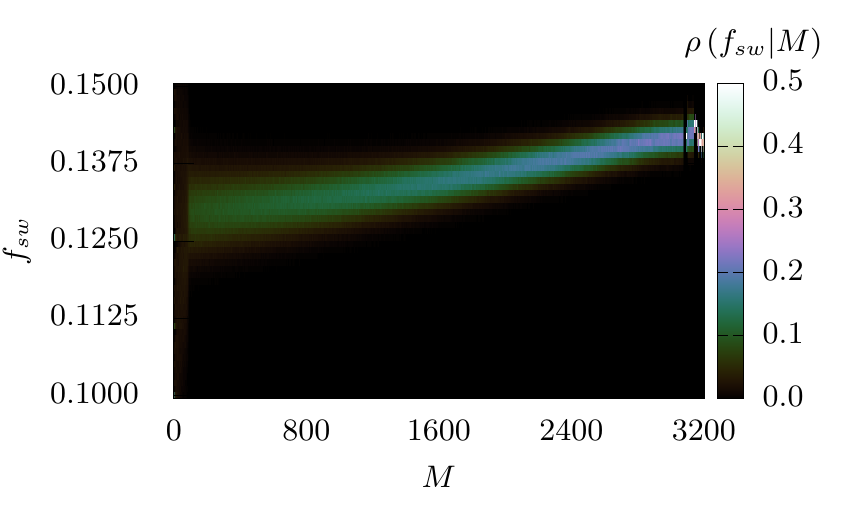}
  \caption{(color online)
Conditional density $\rho\left(f_{\rm sw}|M\right)$, which shows the probability
  of $f_{\rm sw}$ for any given $M$.
  The system size is $N=3200$, the recover probability $\mu=0.14$ and the transmission probability $\lambda=0.1763$}
  \label{max_infected_f_sw}
\end{figure}

The conditional density $\rho\left(f_{\rm sw}|M\right)$ can be found in
\autoref{max_infected_f_sw}.
In contrast to the case when conditioning to $C$, we see a clear
monotonous correlation here: the higher the maximum $M$ of the
$i(\tau)$ time series, the higher the
average fraction of infections which proceeded through long-rang edges.
This makes intuitively sense: As we have seen in \autoref{max_infected_10_to_90_infected} and \autoref{H2_max_infected_min_s}
a higher peak of $i(\tau)$, i.e., higher value of $M$,
is correlated with faster outbreaks.
An infection via long-rang edges should accelerate global spread
and thus the  infection process, leading to larger values of $M$.
The same is observed for $\lambda=0.4$ (not shown). For $\lambda=0.1$, however,
we again see the anti correlation we also observed when conditioning to $C$.
Note that we use the exact same graph for $M$ and $C$ for all three $\lambda$ values.

\section{Summary and outlook}

We investigated outbreak dynamics for diseases described
by the standard SIR model.
Our intention was to investigate typical, extremely mild and extremely
severe outbreaks in principle for arbitrary choices of the transmission
probability $\lambda$
and recovery probability$\mu$. Here we considered a fixed value of $\mu$ and
three representative values of $\lambda$ in the local-outbreak phase,
in the pandemic phase, and near the pandemic threshold $\lambda_c$,
respectively. To achieve this, we
used large-deviation algorithms, in particular a suitably adapted
Wang-Landau approach. We were able to numerically measure, by separate
sets of large-scale simulations, the pdfs of the fraction $C$ of
cumulative infected individuals  and
the peak value $M$ of the fraction of infected
individuals over the whole range of their
support for multiple system sizes.
This allowed us to obtain results
with probability densities as small as $10^{-2500}$.
Furthermore, we were able
to estimate the rate functions for the distributions of these quantities,
showing that the results are compatible with the mathematical
large-deviation property. This means that the SIR process belongs
to a mild or standard class with respect to the large deviations, such
that mathematical tools like the G\"artner-Ellis theorem might be utilized
to obtain analytical progress.

More specifically, we studied networks from the small-world ensemble
for various system sizes up to $N=6400$ nodes.
To gauge our simulation requirements and parameter-space, we first performed simulation of standard SIR dynamics,
 i.e., without the large-deviation approach, to obtain the
critical transmission probability $\lambda_c$ and investigated the
disease duration $\Delta t_{90}$, i.e., how long it takes until 90\% of the
outbreaks were finished.

Beyond obtaining the pdfs,
by comparing the time series that are characteristic for different
regions of the pdfs,
we were able to see three distinct types of outbreaks:
Very mild outbreaks (first region) as well as  very severe pandemic
outbreaks (third region), with respect to
$C$ or $M$ or both, which also evolve very quickly. On the other hand,
outbreaks in the second
region, for intermediate values of $C$ and $M$, behave
somehow chaotic and here we observe the largest times
until the they die down.

In this study, we have investigated the most simple case for the SIR model,
with the intention to provide a case study proving the feasibility of using
large-deviation techniques for epidemic simulations. Clearly, the
approach is not limited to the standard case.
In the future we plan, e.g., to investigate the effect of
disease preventing measures, like
lock downs or government orders to wear masks.
This can be achieved technically by changing the transmission probability
dynamically during an outbreak simulation. The time of change can be
static or depend on the outbreak dynamics.
Any change of the outbreak behavior will be visible
in the measured pdfs, not only in the typical part, but also in the tails,
the structure of the different outbreak dynamics and the measured
correlations. In a similar way, the effect of vaccinations can
in principle easily be measured.

Clearly, the large-deviation
approach is also feasible for extensions of the SIR model,
e.g., when other states are introduced, like \emph{infected but
  not infectious}
or \emph{in quarantine}, or for spatial models, where the mobility plays
a role.
Within a longer perspective, this approach can
also be used to study the rare jump of, e.g., a virus between populations.
This can be achieved by studying two networks simultaneously, i.e.,
a  multilevel network. One network
represents
an animal population, while the second one represents a human population.
Usually
the probability of an animal infecting a human is substantially smaller
than the probability for human-human and animal-animal infections. Thus, such a
transfer leading to a pandemic is a rare event for each single disease.
Hence, this case is ideally suited to be target by a large-deviation approach.

\begin{acknowledgments}
  We thank Peter Werner for critically reading the manuscript.
  
  The simulations were performed at the HPC Cluster CARL, located at the University of Oldenburg (Germany) and funded by the DFG through its
  Major Research Instrumentation Program (INST 184/157-1 FUGG) and the Ministry of Science and Culture (MWK) of the Lower Saxony State.
  We also thank the GWDG G\"ottingen for providing computational resources.
\end{acknowledgments}

\bibliography{sir_paper.bib}

\end{document}